\documentclass[twocolumn,nofootinbib,prd,aps,superscriptaddress,tightenlines]
{revtex4}

\usepackage{amssymb}
\usepackage{bm}
\usepackage{braket}
\usepackage{slashed}
\usepackage{color}
\usepackage{graphicx}
\usepackage{mathrsfs}

{\count255=\time\divide\count255 by 60 \xdef\hourmin{\number\count255}
  \multiply\count255 by-60\advance\count255 by\time
  \xdef\hourmin{\hourmin:\ifnum\count255<10 0\fi\the\count255}}

\newcommand{\nn}{\nonumber \\ }
\newcommand\eUV{\epsilon_{\text{UV}}}
\newcommand\eIR{\epsilon_{\text{IR}}}

\newcommand\lM{\mathsf{L_M}}

\newcommand\lp{\mathsf{L_p}}

\def\cW{c_W}
\def\sW{s_W}
\def\cWsq{c_W^2}
\def\sWsq{s_W^2}

\def\sceth{ $\text{SCET}_{\text{EW}}$}
\def\scetl{ $\text{SCET}_{\gamma}$}

\def\softm{\mathfrak{S}}
\def\softd{\mathfrak{D}}
\def\waver{\mathfrak{R}}

\def\aem{\alpha_{\text{em}}}

\def\msbar{$\overline{\hbox{MS}}$}

\def\vev#1{\left\langle #1 \right\rangle}

\def\rd{{\rm d}}

\begin{document}

\title{Soft and Collinear Functions for the Standard Model}

\author{Jui-yu Chiu}
\affiliation{Department of Physics, University of California at San Diego,
  La Jolla, CA 92093}

\author{Andreas Fuhrer}
\affiliation{Department of Physics, University of California at San Diego,
  La Jolla, CA 92093}

\author{Randall Kelley}
\affiliation{Department of Physics, University of California at San Diego,
  La Jolla, CA 92093}

\author{Aneesh V.~Manohar}
\affiliation{Department of Physics, University of California at San Diego,
  La Jolla, CA 92093}

\begin{abstract}
Radiative corrections to high energy scattering processes were given previously in terms of universal soft and collinear functions. This paper gives the collinear functions for all standard model particles, the general form of the soft function, and explicit expressions for the soft functions for fermion-fermion scattering, longitudinal and transverse gauge boson production, single $W/Z$ production, and associated Higgs production. An interesting subtlety in the use of the Goldstone boson equivalence theorem for longitudinal $W^+$ production is discussed.
\end{abstract}

\date{\today\quad\hourmin}

\maketitle
%\tableofcontents

\section{Introduction}

Hard scattering processes can be described using soft-collinear effective theory (SCET)~\cite{BFL,SCET1,SCET2,BPS}. SCET was extended to broken gauge theories~\cite{CGKM1,CGKM2,CKM,p1} and used to compute the renormalization group improved amplitude for standard model scattering processes at high energy. The effective theory formalism sums the electroweak Sudakov corrections using renormalization group evolution in SCET. The strong and electroweak radiative corrections to hard scattering processes were formulated in terms of collinear and soft functions in Ref.~\cite{p1}. The result gives an efficient way of computing the effective theory radiative corrections in terms of a collinear function for each particle, and universal soft functions.  Electroweak radiative corrections also have been computed previously using fixed-order methods~\cite{ccc,ciafaloni1,ciafaloni2,fadin,kps,fkps,jkps,jkps4,beccaria,dp1,dp2,hori,beenakker,dmp,pozzorini,js,melles1,melles2,melles3,melles4,Denner:2006jr,kuhnW,Denner:2008yn}.

The soft and collinear functions were given in Ref.~\cite{p1} for an $SU(2)$ gauge theory. The complete standard model expressions are more involved because of custodial $SU(2)$ violation, and because the right- and left-handed quarks and leptons have different quantum numbers. In this paper, we give the explicit collinear running and matching functions for each standard model particle, as well as the soft functions for some  important processes such as fermion-fermion scattering, gauge boson pair production, and associated Higgs boson production. We will use the notations and conventions of Ref.~\cite{p1}, and assume that the reader is familiar with the results presented there. The split into soft and collinear contributions is not unique, and we use the definition in Ref.~\cite{p1}. The soft functions for QCD corrections have been obtained previously~\cite{kidonakis}.

A collinear function $\mathscr{F}^{(F\to P)}$ gives the amplitude $F \to P$ for the field $F$ to produce a particle $P$, analogous to the $\braket{0|\phi|p}$ factor in the LSZ reduction formula. Particularly interesting are the collinear functions for $\phi \to W_L$ and $W\to W_T$, $W^3 \to \gamma$, $B\to \gamma$, $W^3\to Z_T$, $B \to Z_T$, and $\phi \to Z_L$ in the Higgs-gauge sector. In most cases, there is a unique $F$, e.g. $u_L$ is only produced by the quark doublet field $Q$, and so the $Q \to u_L$ collinear function is also referred to as the $u_L$ collinear function. The subscript on a fermion field refers to chirality, and on a fermionic particle, to helicity. Thus the $u_R \to u_R$ collinear function is the amplitude for a right-handed $u$ field, with projector $(1+\gamma_5)/2$, to produce a right-handed $u$ quark, with spin parallel to momentum. The difference between helicity and chirality  is order $m/E$, and higher order in the SCET power counting.

We first present plots of the collinear functions in Sec.~\ref{sec:plots} obtained using formul\ae\ given later in Sec.~\ref{sec:col} of the paper. There is an interesting subtlety in the Goldstone boson equivalence theorem for $W^+_L$ arising from infrared divergences due to photon exchange, which is discussed in this section. The general form of the soft functions, and some standard soft matrices are given in Sec~\ref{sec:soft}. These are then used to compute the soft functions for fermion scattering in Sec.~\ref{sec:fermions}, longitudinal and transverse gauge boson production in Sec.~\ref{sec:Wprod}, single-$W,Z$ production in Sec.~\ref{sec:singleW}, and gluon scattering in Sec.~\ref{sec:gluons}.  Appendix~\ref{app:integrate} gives the analytic formula for integrating a SCET anomalous dimension including terms up to the three-loop cusp.

The EFT computation is given by matching from the standard model onto SCET at a scale $\mu_h$, running to $\mu_l$ at which the $W,Z,H,t$ are integrated out, and then running using QCD+QED to a factorization scale $\mu_f$ at which the hadronic scattering cross-sections are computed by convolution with the parton distribution functions. The final answer is independent of the choice of $\mu_{h,l,f}$, but in practice has some dependence on these quantities due to neglected higher order terms. The $\mu_{h,l}$ dependence was shown in Ref.~\cite{p1} to be less than 1\% for processes other than transverse $W_T$ production, for which the $\mu_h$ dependence was almost 10\%.

\section{Plots of Collinear Functions}\label{sec:plots}

In this section, we give numerical plots for the collinear functions for the standard model, and discuss some interesting features of the collinear corrections. The collinear radiative corrections are process independent, and have the same value in all scattering processes.

The collinear functions are given by running the collinear anomalous dimension from $\mu_h$ to $\mu_l$ using the anomalous dimensions in Sec.~\ref{sec:runh}, matching at $\mu_l$ using $\exp D_C$ in Sec.~\ref{sec:matchZ}, and then running from $\mu_l$ to $\mu_f$ using the anomalous dimensions in Sec.~\ref{sec:runl}.  In equations,
\begin{eqnarray}
&&\log \mathscr{F}^{(F \to P)}(\bar n \cdot p,\mu_f,\mu_h)
= -\int_{\mu_f}^{\mu_l} \frac{ {\rm d}\mu}{\mu} \gamma_{P}(\bar n \cdot p,\mu)\nn
&&+ D_C^{(F \to P)}(\bar n \cdot p,\mu_l) -\int_{\mu_l}^{\mu_h} \frac{ {\rm d}\mu}{\mu} \gamma_{F}(\bar n \cdot p,\mu)\,.
\end{eqnarray}
The collinear corrections are functions of $\bar n \cdot p =2E$, where $E$ is the particle energy,\footnote{The collinear functions depend on the Lorentz frame through the null-vector $n$. The $n$ dependence is cancelled by a corresponding $n$ dependence in the soft functions, by reparametrization invariance~\cite{rpi1,rpi2}.} and depend linearly on $\log \bar n \cdot p$ to all orders in perturbation theory~\cite{dis}. They are defined after zero-bin subtraction to avoid double counting with the soft contribution~\cite{hautmann,zerobin,leesterman,idilbimehen1,idilbimehen2,Delta,Chiu:2009yz,messenger}. These subtractions are necessary for soft-collinear factorization~\cite{HardScattering} to hold. 

The collinear functions were used to compute $2 \to 2$ scattering processes in Ref.~\cite{p1}, where we used $\mu_h = \sqrt{s_0}$ for the high-scale matching. In the partonic center-of-mass frame, all four partons have energy $2E=\bar n \cdot p=\sqrt{s_0}$.  For this reason, we have used $\mu_h=\bar n \cdot p$ in the collinear function plots, to reduce the number of variables by one. The low scales are chosen to be $\mu_f=\mu_l=M_Z$. There is a tiny dependence on the Higgs mass---the rates change by less than one part in $10^4$ if $m_H$ is varied between 200 and 500~GeV. In the plots, $m_H=200$~GeV. The collinear anomalous dimension can be integrated analytically using the results in Appendix~\ref{app:integrate}.  If the factorization scale $\mu_f$ is below $M_Z$, there is an additional contribution to the collinear function from QCD+QED running from $M_Z$ to $\mu_f$, which is given separately. The total collinear function is the product of the $\mu_h \to M_Z$ and $M_Z \to \mu_f$
collinear functions.

\begin{figure}
\begin{center}
\includegraphics[bb=60 152 473 713,width=9cm]{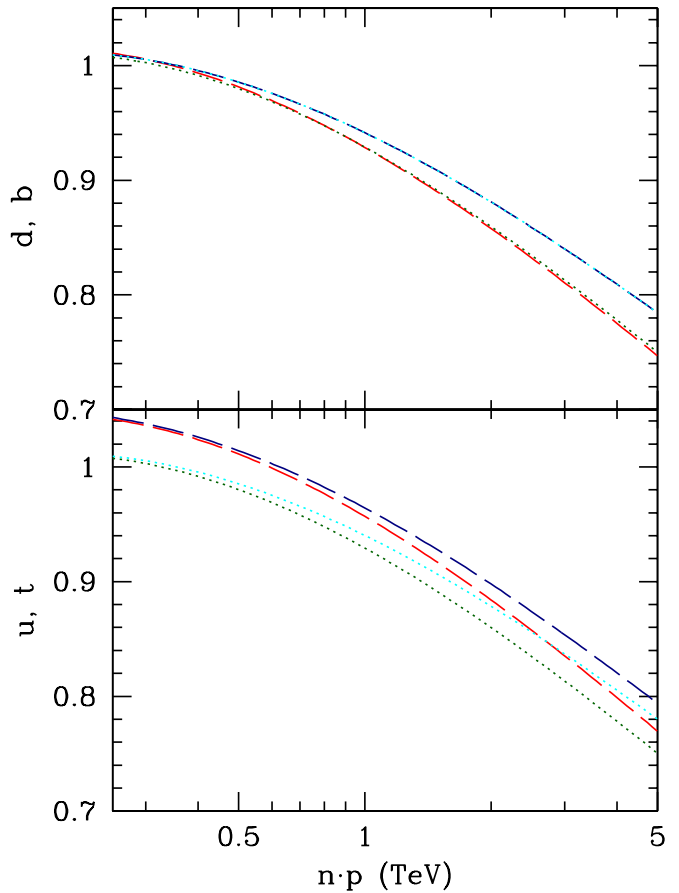} 
\end{center}
\caption{\label{fig:colQ}Plot of the collinear functions against $\bar n \cdot p$ for (a) lower panel: $u_L$ (dotted green), $u_R$ (dotted cyan), $t_L$ (dashed red), $t_R$ (dashed blue) (b) upper panel:
$d_L$ (dotted green), $d_R$ (dotted cyan), $b_L$ (dashed red), $b_R$ (dashed blue). }
\end{figure}

Fig.~\ref{fig:colQ} 
%%%%%%%%%%%%%%%%%%%%%%%%%
%%%%%%%%%%%%%%%%%%%%%%%%%
shows the collinear functions for quarks. The  collinear functions for the $c$ and $s$ are identical to those for the $u$ and $d$, respectively. The $t$ and $b$ quarks have slightly different collinear functions because of Higgs corrections, and the mass of the $t$. In a $2 \to 2$ scattering process such as $u_L \bar u_L \to d_L \bar d_L$, one has a collinear function  in the amplitude for each external particle, so the rate depends on the product of the fourth powers of the $u_L$ and $d_L$ collinear functions. Thus a 10\% correction in Fig.~\ref{fig:colQ} changes the rate by more than a factor of two. The difference between the heavy- and light-quark collinear functions arises from Higgs contributions due to the $t_R$ Yukawa coupling to the quark doublet $Q^{(t)}$, and due to the switch from SCET to bHQET fields for the $t$.

%%%%%%%%%%%%%%%%%%%%%%%%%
\begin{figure}
\begin{center}
\includegraphics[bb=60 152 473 470,width=9cm]{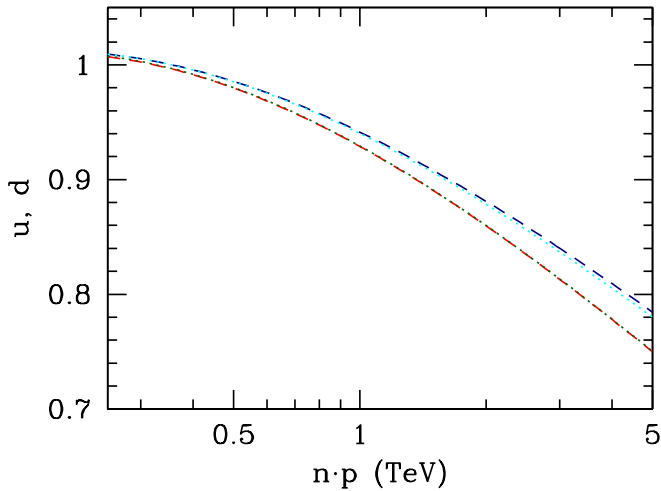} 
\end{center}
\caption{\label{fig:colQl}Plot of the collinear functions against $\bar n \cdot p$ for $u_L$ (dotted green), $u_R$ (dotted cyan), $d_L$ (dashed red), and $d_R$ (dashed blue). }
\end{figure}
%%%%%%%%%%%%%%%%%%%%%%%%%

\begin{figure}
\begin{center}
\includegraphics[bb=60 152 473 470,width=9cm]{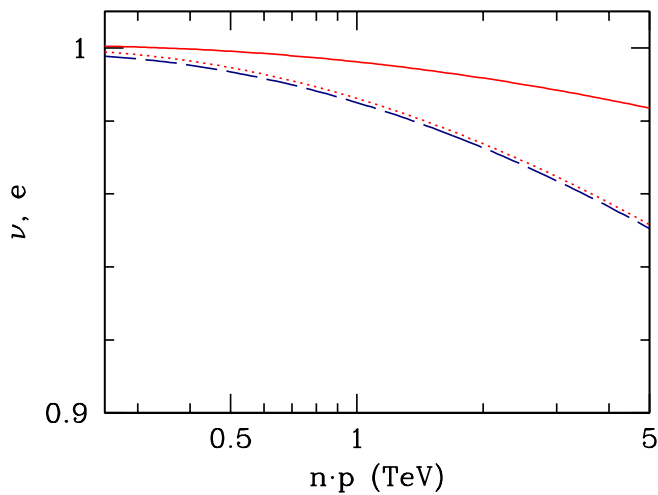} 
\end{center}
\caption{\label{fig:colE}Plot of the collinear functions against $\bar n \cdot p$ for $\nu_L$ (dashed blue), $e_L$ (dotted red),  and $e_R$ (solid red). }
\end{figure}

Fig.~\ref{fig:colQl} shows the collinear functions for $u$ and $d$ on the same plot. The left- and right-handed quarks have different collinear functions because of the difference in $SU(2)$ quantum numbers. There is a small difference between $u_R,d_R$ due to the different $U(1)$ quantum numbers, which lead to different $U(1)$ anomalous dimensions. There is an even smaller difference between $u_L,d_L$ due to differences in the low-scale matching from $Z$ exchange due to the different $Z$ couplings. Fig.~\ref{fig:colE} shows the collinear functions for the leptons. The corrections are smaller than for quarks because there are no QCD corrections.
%%%%%%%%%%%%%%%%%%%%%%%%%
%%%%%%%%%%%%%%%%%%%%%%%%%

\begin{figure}
\begin{center}
\includegraphics[bb=60 152 473 713,width=9cm]{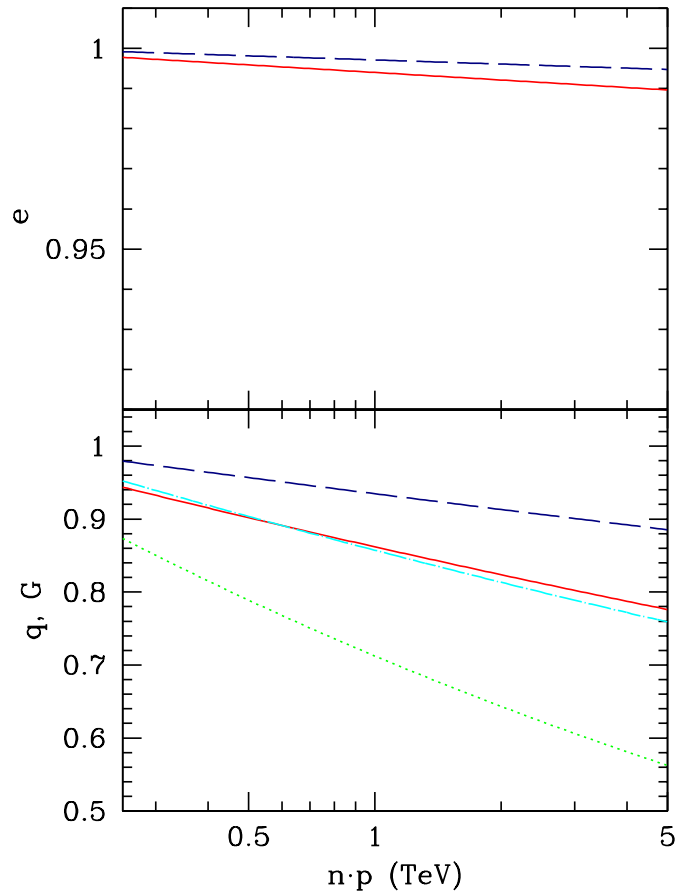} 
\end{center}
\caption{\label{fig:colLow}Plot of the collinear functions due to running from $M_Z$ to $\mu_f$ against $\bar n \cdot p$ for
electrons with $\mu_f=30$~GeV (solid red) and $\mu_f=50$~GeV (dashed blue) are shown in the upper panel. The QCD correction for quarks with $\mu_f=30$~GeV (solid red) and $\mu_f=50$~GeV (dashed blue) and gluons with $\mu_f=30$~GeV (dotted green) and $\mu_f=50$~GeV (dot-dashed cyan) are shown in the lower panel. }
\end{figure}

If the factorization scale is chosen below $M_Z$, there is additional collinear running from QCD and QED. The QCD collinear running is the same for all quarks, and the log of the QED running is proportional to the electric charge. Fig.~\ref{fig:colLow} show the collinear running below $M_Z$ for $\mu_f=30,50$~GeV for quarks, gluons and electrons. These multiply the collinear running from $\mu_h$ to $M_Z$.
%%%%%%%%%%%%%%%%%%%%%%%%%
%%%%%%%%%%

%%%%%%%%%%%%%%%%%%%%%%%%%
\begin{figure}
\begin{center}
\includegraphics[bb=60 152 473 470,width=9cm]{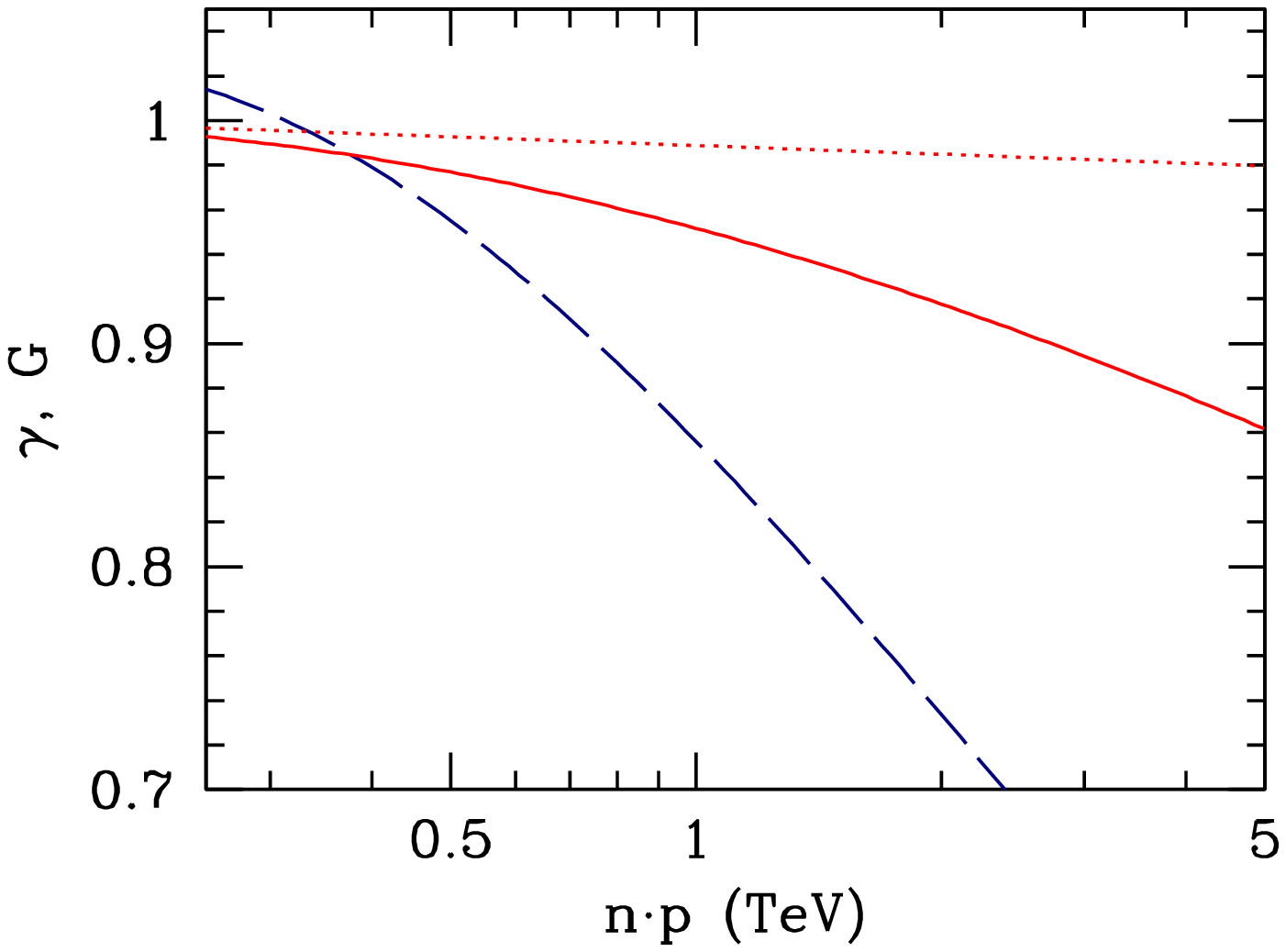} 
\end{center}
\caption{\label{fig:colG}Plot of the collinear functions against $\bar n \cdot p$ for $W \to \gamma$ (solid red), $B \to \gamma$ (dotted red) and gluons (dashed blue).}
\end{figure}
%%%%%%%%%%

The collinear functions for massless gauge bosons are shown in Fig.~\ref{fig:colG}. The corrections to the gluon are due to QCD, and are large because of the large value of $C_A$.
There are two collinear functions for photon production, depending on the source of the photon. The $W^3-B$ and $Z-\gamma$ fields are related by
\begin{eqnarray}
Z &=& \cos \theta_W W^3-\sin\theta_W B\,,\nn
A &=& \sin \theta_W W^3+\cos\theta_W B\,.
\end{eqnarray}
At tree-level the $W^3 \to \gamma$ amplitudes is $\sin \theta_W$, and the $B \to \gamma$ amplitude is $\cos \theta_W$. The photon can be emitted by what started out as either a $W^3$ or $B$ field at high energy, and the radiative corrections shown by the solid red and dotted red curves in Fig.~\ref{fig:colG} multiply the tree-level amplitudes. The correction for $W\to \gamma$ is much larger because of the $SU(2)$ contribution.
 
%%%%%%%%%%%%%%%%%%%%%%%%%
\begin{figure}
\begin{center}
\includegraphics[bb=60 152 473 713,width=9cm]{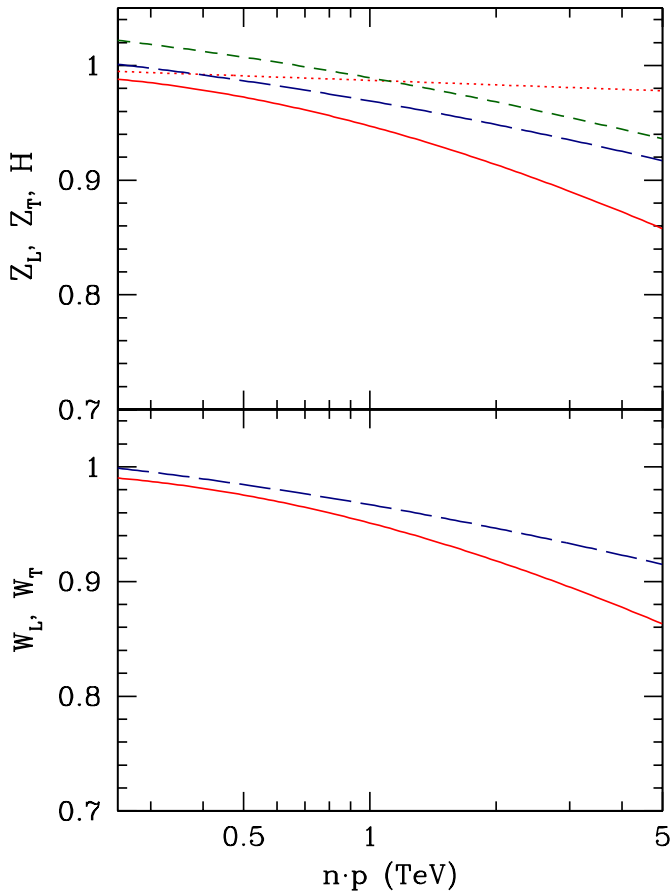} 
\end{center}
\caption{\label{fig:colW}Plot of the collinear functions against $\bar n \cdot p$ for (a) lower panel: $W_T$ (solid red), $W_L$ (dashed blue) (b) upper panel:
$W \to Z_T$ (solid red), $B \to Z_T$ (dotted red), $Z_L$ (dashed blue) and $H$ (short-dash, dark green).}
\end{figure}
%%%%%%%%%%%%%%%%%%%%%%%%%
Fig.~\ref{fig:colW} gives the collinear functions for the massive gauge bosons and Higgs. The lower panel shows the collinear functions for the transverse and longitudinal $W$, i.e. for $W \to W_T$ and $\phi \to W_L$, since the transverse $W$ can only come from the $W$ field and the longitudinal $W$ from the $\phi$ field. 
The radiative corrections are different, because at high energies, $W_T$ is part of the $W_\mu$ gauge field, whereas $W_L$ is part of the scalar field $\phi$. The $SU(2)$ corrections to $W_\mu$ depend on the adjoint Casimir $C_A=2$, whereas the corrections to $\phi$ depend on the fundamental Casimir $C_F=3/4$. The $U(1)$ corrections also differ. At high energies, the $W_L$ remembers that it originated from the scalar field via spontaneous symmetry breaking. The upper panel gives the plots for the neutral boson sector. The transverse $Z$ can arise from either $W^3$ or $B$, as for the photon, and the two cases are shown in solid and dotted red. The $B \to Z_T$ amplitude has smaller corrections (as for $ B \to \gamma$), so at high energies, $Z_T$ is produced mainly via $B \to Z_T$, even though at tree-level, it is the $W^3 \to Z$ amplitude which dominates. The $\phi \to Z_L$ and $\phi \to H$ amplitudes have similar shapes, since both are mainly given by the radiative corrections to the scalar doublet $\phi$. There are two amplitudes for $Z_T$, $W \to Z$ and $B \to Z$, but only one for $Z_L$, $\phi \to Z$.

\section{Collinear Functions}\label{sec:col}

\begin{figure}
\begin{center}
\includegraphics[width=6cm]{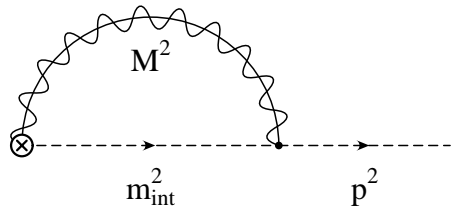} 
\end{center}
\caption{\label{fig:fd7} One loop collinear graph, where the  internal and external 
particles can have different masses, e.g. $m_{\text{int}}=m_b$ and $p^2=m_t^2$. The wavy+solid line is the collinear gauge boson, and the dashed line is a collinear fermion or scalar.}
\end{figure}

The formul\ae\ for the collinear functions are given in this section. They were obtained using the procedure given in Ref.~\cite{p1}. The main complication arises from custodial $SU(2)$ symmetry breaking in the standard model. In loop graphs, one has to distinguish between $W$ and $Z$ exchange as well as the $m_t$-$m_b$ mass difference. The collinear functions, computed from one-loop graphs such as Fig.~\ref{fig:fd7} are summarized in Table~\ref{tab:massless}. The anomalous dimension $\gamma_C$ gives the running between the high-scale $\mu_h \sim \sqrt{\hat s}$ and the low-scale $\mu_l \sim M_Z$, and the matching $D_C$ gives the collinear matching at the low scale $\mu_l$. The $\gamma_C$ column can also be used to obtain the anomalous dimension in \scetl\  between $\mu_l$ and the factorization scale $\mu_f$.
\begin{table*}
\begin{eqnarray*}
\renewcommand{\arraystretch}{1.5} 
\begin{array}{c|c|c|}
\text{Field} & \gamma_C & D_C \\
\hline
\psi & \frac{\alpha }{4\pi}\mathbf{T}\cdot\mathbf{T}\left[ 4 \lp - 4\right] +\gamma_\psi & \frac{\alpha }{4\pi}\mathbf{T}\cdot\mathbf{T}\left[2 \lM \lp - \frac12 \lM^2-2 \lM - \frac{5\pi^2}{12} + 
2+f_F(p^2/M^2,m_{\text{int}}^2/M^2)\right] +\frac12 \delta \waver_\psi\\
\phi &  \frac{\alpha }{4\pi}\mathbf{T}\cdot\mathbf{T}\left[ 4\lp - 2 \right] +\gamma_\phi &  \frac{\alpha }{4\pi}\mathbf{T}\cdot\mathbf{T}\left[ 2 \lM \lp - \frac12 \lM^2- \lM - \frac{5\pi^2}{12} + 1+f_S(p^2/
M^2,m_{\text{int}}^2/M^2)\right] +\frac12 \delta \waver_\phi\\
h_v & \frac{\alpha }{4\pi}\mathbf{T}\cdot\mathbf{T}\left[ 4\log (2\gamma) \right]+\gamma_h & \frac{\alpha }{4\pi}\mathbf{T}\cdot\mathbf{T}\left[ 2\lM \log2\gamma \right] +\frac12 \delta \waver_{h_v}\\
B_\perp  &  \frac{\alpha }{4\pi}\mathbf{T}\cdot\mathbf{T}\left[4 \lp - 2\right]+\gamma_W &   \frac{\alpha }{4\pi}\mathbf{T}\cdot\mathbf{T}\left[ 2 \lM \lp - \frac12 \lM^2- \lM - \frac{5\pi^2}{12} + 
1 +
f_S(1,1)\right] +\frac12 \delta \waver_W\\
H &  \frac{\alpha }{4\pi}\mathbf{T}\cdot\mathbf{T}\left[4 \lp - 2\right] + \gamma_H &  \frac{\alpha }{4\pi}\mathbf{T}\cdot\mathbf{T}\left[2 \lM \lp - \frac12 \lM^2- \lM - \frac{5\pi^2}{12} + 1
+f_S(m_h^2/M^2,1)\right]+\frac12 \delta \waver_H\\
\varphi^a &  \frac{\alpha }{4\pi}\mathbf{T}\cdot\mathbf{T}\left[4 \lp-2\right]+\gamma_\varphi &  \frac{\alpha }{4\pi}\mathbf{T}\cdot\mathbf{T}\left[2 \lM \lp - \frac12 \lM^2- \lM - \frac{5\pi^2}{12} 
+ 1
+\frac23 f_S(1,1)+\frac13 f_S(1,m_h^2/M^2)\right]+\frac12 \delta \waver_\varphi\\
\hline
\end{array}
\end{eqnarray*}
\caption{\label{tab:massless} The collinear anomalous dimension and low-scale matching.  $\lM=\log(M^2/\mu^2)$, $\lp=\log (\bar n \cdot p)/\mu$, and $\gamma=E/m$. The rows are $\psi$: fermion, $\phi$ non-Higgs scalar multiplet, $h_v$ HQET field, $B_\perp$: transverse gauge boson, $H$: Higgs, $\varphi^a$: Goldstone bosons (i.e.\ longitudinal gauge bosons using the equivalence theorem and mutiplying by $\mathcal{E}$). The results are in $R_{\xi=1}$ gauge. $\gamma_{W,h,\varphi}$ and $\waver_{W,h,\varphi}$ are the 
wavefunction contributions. $p^2$ is $m^2$ for the external particle, and $m_{\text{int}}$ is the mass of the internal particle.}
\end{table*}
%%%--FIGURE--------------------------------------------------------------------------------------
%%%-------------------------------------------------------------------------------------------------
This table is a generalization of Table~II of Ref.~\cite{p1}, which gave the collinear functions in the $SU(2)$ theory.  In the  weak interactions, the two members of an $SU(2)$ doublet can have different masses. As a result, in computing Fig.~\ref{fig:fd7}, the internal and external fermions can have different  masses; e.g.\ the internal fermion can be a $b$-quark, and the external one, a $t$-quark. This complication did not arise for the $SU(2)$ theory with massless fermions considered in Ref.~\cite{p1}. The collinear functions in Table~\ref{tab:massless} include the possibility of different internal and external masses. $m_{\text{int}}$ is the mass of the internal particle in the loop, and  $\sqrt{p^2}$ is the mass of the external particle. The functions $f_{F,S}$ are given in Appendix~B of Ref.~\cite{CKM}, and vanish for massless particles, $f_F(0,0)=f_S(0,0)=0$. The $\psi$ row is for fermions, $\phi$ for scalars, $B_\perp$ for an external  transversely polarized gauge boson, $H$ for the physical Higgs field, and $\varphi^a$ for the Goldstone bosons, which are used to compute longitudinally polarized gauge bosons using the equivalence theorem.

Table~\ref{tab:massless} gives the results in a compressed form, from which the standard model results can be extracted.
The $\mathbf{T}\cdot \mathbf{T}$ factor has to be taken apart into individual gauge boson contributions
\begin{eqnarray}
\alpha\ \mathbf{T}\cdot\mathbf{T} &\to&  \alpha_s \mathbf{T}\cdot\mathbf{T}
+\alpha_2 \mathbf{t}\cdot \mathbf{t}  + \alpha_1 Y\cdot Y\,,
\label{rewrite}
\end{eqnarray}
summing over the $SU(3)$, $SU(2)$ and $U(1)$ contributions, where $\mathbf{T}$ are the QCD generators, and $\mathbf{t}$ are the $SU(2)$ generators. This form is convenient for computing the collinear anomalous dimension $\gamma_C$, which is mass-independent. The electroweak couplings constants are $\alpha_2=\aem/\sin^2\theta_W$ and $\alpha_1=\aem/\cos^2\theta_W$.

The low-scale collinear matching $D_C$ depends on the gauge boson masses, so  the $SU(2) \times U(1)$ part has to be rewritten in terms of the $W$, $Z$ and $\gamma$ contributions,
\begin{eqnarray}
\alpha_2 \mathbf{t}\cdot \mathbf{t}  + \alpha_1 Y\cdot Y &\to&
 \frac12 \alpha_W \left(t_+ t_- + t_- t_+ \right)\nn
 && + \alpha_Z t_Z \cdot t_Z+ \aem Q\cdot Q
 \label{su2u1}
 \end{eqnarray}
where $\alpha_W=\alpha_2$ and $\alpha_Z=\aem/(\sin^2\theta_W \cos^2\theta_W)$ and $t_Z$ is the $Z$-charge, $t_Z=t_3-\sin^2\theta_W Q$. A useful identity for the $W$ contribution is
 \begin{eqnarray}
\frac12\left(t_+ t_- + t_- t_+\right) &=&  \mathbf{t}\cdot \mathbf{t}  - t_3\cdot t_3\,.
 \label{su2u1a}
\end{eqnarray}

The matching $D_C$ depends on the gauge boson and fermion masses. The value of  $D_C$ is given using Table~\ref{tab:massless} and Eqs.~(\ref{rewrite},\ref{su2u1}) and  using $M \to M_W$ in the $W$ terms and $M \to M_Z$ in the $Z$ terms. The photon and gluon do not contribute to $D_C$, since they are not integrated out at the low-scale $\mu_l \sim M_Z$, and are dropped. Furthermore, in $f_F$ and $h_F$, the internal fermion mass is equal to the external fermion mass for the $Z$ term, but is different for the $W$ term. For example, for an external $t$ quark, $p\to m_t$, $m_{\text{int}}\to m_t$ for the $t_Z^2$ term and $p\to m_t$, $m_{\text{int}}\to m_b$ for the $t_+t_-$ and $t_-t_+$ terms. Explicit formul\ae\ for the standard model particles are given below using this procedure.

The wavefunction factors $\delta\waver_{\varphi,H,W}$ can be found in Ref.~\cite{bohmbook,bardin,fleischer,hollik}. 
They are defined as the residue of the two-point Green's function at the pole,
\begin{eqnarray}
G &\sim& \frac{ \waver}{p^2-M^2} + \text{finite}\,,
\end{eqnarray}
with $\waver=1+\delta \waver$. $\waver$ is obtained using the two-point function renormalized in the $\overline{\text{MS}}$ scheme, and is finite. We use the convention of Ref.~\cite{bohmbook} and denote the finite wavefunction correction by $\waver$, and reserve $Z$ for the infinite renormalization counterterms. There is one important point to remember --- the wavefunction graphs have to be computed as an EFT matching condition. This means that the graphs are computed using dimensional regularization to regulate the infrared divergences, setting all low energy scales such as $m_b$ to zero, and retaining only the finite part.\footnote{\label{foot} See, for example, Refs.~\cite{amhqet,ameft,dis} for a more extensive discussion and explicit examples. In Eqs.~(\ref{eq2}-\ref{eq9}), the subscripts UV and IR indicate whether the divergence is ultraviolet or infrared. The integrals are done in $4-2\epsilon$ dimensions, so $\eUV=\eIR=\epsilon$.} $\waver$ can be obtained from the expressions in terms of Passarino-Veltman functions using
\begin{eqnarray}
A_0(m^2) &=& -m^2 \left(\frac{1}{\eUV}+1- \ln \frac{m^2}{\mu^2}\right)\,,
\label{eq2}
\end{eqnarray}
and
\begin{eqnarray}
&& B_0(p^2,m_1,m_2) \nn
&&= \Gamma(\epsilon)e^{\epsilon \gamma}\mu^{2\epsilon}
\int_0^1{\rm d}x\ \left[m_1^2x+m_2^2(1-x)+p^2 x (1-x)\right]^{-\epsilon}\,,\nn
&&B_0^\prime(-M^2,m_1,m_2) = \left.\frac{\partial B_0(p^2,m_1,m_2)}{\partial p^2}\right|_{p^2=-M^2} 
\label{B0def}
\end{eqnarray}
where we follow the conventions of Ref.~\cite{bardin}. In particular, the infrared divergent functions needed are
\begin{eqnarray}
B_0(0,0,0) &=& \frac{1}{\eUV}-\frac{1}{\eIR}
\end{eqnarray}
and
\begin{eqnarray}
B_0^\prime(-M^2,0,M)
&=&\frac{1}{M^2}\left[\frac{1}{2\eIR}+1 -\frac12 \log \frac{M^2}{\mu^2}\right]\,,\nn
\label{eq9}
\end{eqnarray}
which are replaced by $0$ and $(1-1/2 \log M^2/\mu^2)/M^2$, respectively, in $\waver$.

In Refs.~\cite{CGKM2,CKM},  the radiative corrections for massive particles were computed. In the region below the particle mass, the particle can be  treated as a bHQET field~\cite{top1,top2}.  The anomalous dimension and low-scale matching for bHQET fields is  given in the row $h_v$. For massive particles, the collinear anomalous dimension involves $\log 2 \gamma$, where $\gamma=E/m$ is the boost factor, rather than $\log (\bar n \cdot p)/\mu=\log(2E)/\mu$. The bHQET formula is needed for top-quark pair production, and for $W$ and $Z$ production.

In addition to gauge boson exchange, there are radiative corrections due to scalar exchange graphs.  In the standard model, these arise from Higgs exchange. As shown in Refs.~\cite{CGKM2,CKM}, scalar exchange vertex graphs are $1/Q^2$ suppressed, and only the wavefunction graphs are leading order in the SCET power counting.  Thus we can  include Higgs corrections in the effective theory through their contribution to $\waver$.

The anomalous dimension between $\mu_h$ and $\mu_l$ is independent of the low-energy scales, including the electroweak symmetry breaking scale, and so can be computed in the unbroken gauge theory. The collinear functions depend on $\bar n \cdot p = 2E$, where $E$ is the energy of the particle. The $\bar n$ dependence, or Lorentz frame dependence, is cancelled by a corresponding frame dependence in the soft functions.

The left-handed quark doublets will be denoted by $Q^{(i)}_L$, where $i=u,c,t$ is a flavor index, the right-handed charge $2/3$ quarks by $U^{(i)}_R$ or $u_R,c_R,t_R$, the right-handed charge $-1/3$ quarks by $D^{(i)}_R$, or $d_R,s_R,b_R$, the left-handed lepton doublets by  $L^{(i)}_L$, $i=e,\mu,\tau$, and the right-handed lepton singlets by  $E^{(i)}_R$ or $e_R,\mu_R,\tau_R$. Written in terms of $SU(2)$ components, $Q^{(i)}$ is
%----------------------------------------------------------------
\begin{eqnarray}
Q^{(i)} = \left[ \begin{array}{cc} U^{(i)}_L \\[5pt] D_L^{\prime(i)} \end{array}\right]=
\left[ \begin{array}{cc} U^{(i)}_L \\[5pt] V_{ij} D_L^{(j)} \end{array}\right]\,,
\end{eqnarray}
where the primed down-type quarks are weak eigenstate fields, and the unprimed fields are mass eigenstates. All the lepton and down-type quark masses can be neglected in our calculation, so we can work in the weak eigenstate basis, the CKM matrix $V$ does not enter the SCET computation, and generation number is conserved. Once the radiative corrections have been computed, one can make the replacement $D_L^{\prime(i)} \to V_{ij} D_L^{(j)}$ to compute the amplitudes in terms of mass-eigenstate fields.

\subsection{Running from $\mu_h$ to $\mu_l \sim M_Z$}\label{sec:runh}

The collinear anomalous dimensions for the running from $\mu_h \sim Q$ to $\mu_l \sim M_Z$ are listed below. The gauge coupling constants are the \msbar\ values in the theory with six dynamical quark flavors, and $y_t = \sqrt{2} m_t/v$ is the $t$-quark Yukawa coupling. The top quark multiplets have different collinear running than the other quarks because of the large Yukawa coupling $y_t$.

$Q^{(u,c)}_L$:
\begin{eqnarray}
\frac{1}{4\pi}\left(\frac43 \alpha_s+\frac 3 4\alpha_2+\frac 1 {36} \alpha_1 \right)\left(4 \log \frac{\bar n \cdot p}{\mu} - 3 \right)
\end{eqnarray}

$Q^{(t)}_L$:
\begin{eqnarray}
\frac{1}{4\pi}\left(\frac43 \alpha_s+\frac 3 4\alpha_2+\frac 1 {36} \alpha_1 \right)\left(4 \log \frac{\bar n \cdot p}{\mu} - 3 \right)
+\frac12\frac{y_t^2}{16\pi^2} \nn
\end{eqnarray}

$u_R,\ c_R$:
\begin{eqnarray}
\frac{1}{4\pi}\left(\frac43 \alpha_s+\frac {4} {9}\alpha_1  \right)\left(4 \log \frac{\bar n \cdot p}{\mu} - 3 \right)
\end{eqnarray}

$t_R$:
\begin{eqnarray}
\frac{1}{4\pi}\left(\frac43 \alpha_s+\frac {4} {9}\alpha_1  \right)\left(4 \log \frac{\bar n \cdot p}{\mu} - 3 \right)+\frac{y_t^2}{16\pi^2}
\end{eqnarray}

$d_R,s_R,b_R$:
\begin{eqnarray}
\frac{1}{4\pi}\left(\frac43 \alpha_s+\frac {1} {9} \alpha_1 \right)\left(4 \log \frac{\bar n \cdot p}{\mu} - 3 \right)
\end{eqnarray}

$L^{(e)}_L,\ L^{(\mu)}_L,\ L^{(\tau)}_L$:
\begin{eqnarray}
\frac{1}{4\pi}\left(\frac 3 4 \alpha_2+\frac 1 {4}\alpha_1  \right)\left(4 \log \frac{\bar n \cdot p}{\mu} - 3 \right)
\end{eqnarray}

$e_R,\ \mu_R,\ \tau_R$:
\begin{eqnarray}
\frac{\alpha_1}{4\pi} \left(4 \log \frac{\bar n \cdot p}{\mu} - 3 \right)
\end{eqnarray}

The gauge field anomalous dimension at one-loop in $R_{\xi=1}$ gauge is
\begin{eqnarray}
\gamma &=& 2 C_A - b_0\,,
\end{eqnarray}
where $b_0$ is the coefficient of the first term in the $\beta$-function,
\begin{eqnarray}
\mu \frac{\rd g}{\rd \mu} &=& -b_0 \frac{g^3}{16\pi^2}+\ldots
\end{eqnarray}
so that the collinear factor $\gamma_C$ for transverse gauge bosons is
\begin{eqnarray}
\frac{\alpha }{4\pi}\left(4 C_A \lp - b_0\right)\,.
\end{eqnarray}
It is more convenient to write the anomalous dimensions for $W_3$ and $B$ instead of $Z$ and $\gamma$, to avoid off-diagonal mixing terms in the renormalization group evolution due to the running of $\sin^2\theta_W$.

$g$ (transverse gluons):
\begin{eqnarray}
\frac{\alpha_s}{4\pi}\left(12 \log \frac{\bar n \cdot p}{\mu} -7\right)
\end{eqnarray}

$W_T$ (transverse $W^{1,2,3}$):
\begin{eqnarray}
\frac{\alpha_2}{4\pi}\left(8 \log \frac{\bar n \cdot p}{\mu} -\frac{19}{6}\right)
\end{eqnarray}

$B_T$ (transverse $B$):
\begin{eqnarray}
\frac{\alpha_1}{4\pi}\left(\frac{41}{6}\right)\,.
\end{eqnarray}

The scalar anomalous dimension for the unphysical Goldstone bosons, needed for longitudinal gauge boson production using the equivalence theorem, and for Higgs production is\\
$\varphi, H$:
\begin{eqnarray}
\frac{1}{4\pi}\left(\frac34 \alpha_2+\frac {1} {4} \alpha_1 \right)\left(4 \log \frac{\bar n \cdot p}{\mu} -4\right)+3 \frac{y_t^2}{16\pi^2}\,.
\end{eqnarray}
The $y_t$ term in the $\phi$ anomalous dimension affects the rates for $H$, $W_L$ and $Z_L$ production at the few percent level.

\subsection{Matching at $\mu_l \sim M_Z$}\label{sec:matchZ}

The matching corrections at $\mu_l \sim M_Z$ have to be computed in the broken electroweak theory, using Table~\ref{tab:massless}, Eq.~(\ref{su2u1}) and the discussion following it. The matching can be computed for each particle, and is shown schematically
%%%----FIGURE--------------------------------------------------------------------------------------
\begin{figure}
\begin{center}
\includegraphics[width=4cm]{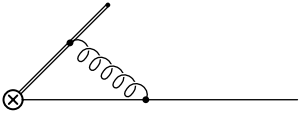} 
\end{center}
\caption{\label{fig:colmatching}Collinear matching graphs for $[W^\dagger \psi]$. The $\otimes$ is the $[W^\dagger \psi]$ operator, the solid line is $\psi$ and the double line is $W^\dagger$.}
\end{figure}
in Fig.~\ref{fig:colmatching}. The collinear gauge invariant operator $[W^\dagger_{\text{EW}} \psi]$ in \sceth\ matches onto $[W^\dagger_\gamma \psi]$ in \scetl. The difference between the collinear Wilson lines is that $W^\dagger_{\text{EW}}$ contains gluons, $W$ and $B$ gauge fields which are the dynamical fields in \sceth\, whereas $W^\dagger_\gamma$ contains gluons and photons, the dynamical gauge fields in \scetl. The matching coefficients are given by integrating out the $W$ and $Z$. Once again, the collinear matching is more complicated due to custodial $SU(2)$ violation. Thus, in the quark doublet $Q_L=(u,d)_L$, there are separate matching functions for $u_L$ and $d_L$, etc.

The low-scale matching for the quark doublet can be written as:
\begin{eqnarray}
\left[W^\dagger_{\text{EW}} Q_L\right] & \rightarrow & \left[ \begin{array}{c} 
\exp D_C^{(Q_L \to U_L)}\ [W^\dagger_\gamma U_L] \\
\exp D_C^{(Q_L \to D_L)}\ [W^\dagger_\gamma D_L] \\
\end{array}\right]\,.
\label{qmatch}
\end{eqnarray}
The quantity $\left[W^\dagger_{\text{EW}} Q_L\right]_a$ is collinear gauge-invariant, and has an index $a$. Eq.~(\ref{qmatch}) implies that the $a=1$ term matches to $[W^\dagger_\gamma U_L]$ and the $a=2$ term to $[W^\dagger_\gamma D_L]$, with amplitudes $\exp D_C^{(Q_L \to U_L)}$ and $\exp D_C^{(Q_L \to D_L)}$, respectively.  The other cases listed below use a similar notational convention. The collinear functions $D_C$ are zero at tree-level.

The remaining fermionic collinear matching functions are defined by:
\begin{eqnarray}
\left[W^\dagger_{\text{EW}} U_R \right] & \rightarrow & \exp D_C^{(U_R \to U_R)}\  [W^\dagger_\gamma U_R]\,, \nn
\left[W^\dagger_{\text{EW}} D_R\right] & \rightarrow & \exp D_C^{(D_R \to D_R)}\  [W^\dagger_\gamma  D_R] \,,\nn
\left[W^\dagger_{\text{EW}} L_L\right] & \rightarrow & \left[ \begin{array}{c} 
\exp D_C^{(L_L \to \nu_L)}\ \nu_L \\
\exp D_C^{(L_L \to E_L)}\  [W^\dagger_\gamma  E_L] \\
\end{array}\right]\,,\nn
\left[W^\dagger_{\text{EW}} E_R\right] & \rightarrow & \exp D_C^{(E_R \to E_R)}\  [W^\dagger_\gamma  E_R]\,.
\end{eqnarray}

The collinear matching functions are:
\begin{eqnarray}
D_C^{(Q_L \to U_L)}(\mu)&=&
g_{LU}^2 D_Z(\mu)+\frac12D_W(\mu)\,,\nn
D_C^{(Q_L \to t_L)}(\mu)&=&
g_{LU}^2 D_Z(\mu)+\frac12D_W(\mu)+F_{t_L}(\mu)\,,\nn
D_C^{(Q_L \to D_L)}(\mu)&=&
g_{LD}^2 D_Z(\mu)+\frac12D_W(\mu)\,, \nn
D_C^{(Q_L \to b^\prime)}(\mu)&=&
g_{LD}^2 D_Z(\mu)+\frac12D_W(\mu)+F_{b^\prime_L}(\mu) \,,\nn
D_C^{(U_R \to U_R)}(\mu)&=&
g_{RU}^2 D_Z(\mu)\,,\nn
D_C^{(t_R \to t_R)}(\mu)&=&
g_{RU}^2 D_Z(\mu)+F_{t_R}(\mu)\,,\nn
D_C^{(D_R \to D_R)}(\mu)&=&
g_{RD}^2 D_Z(\mu)\,,\nn
D_C^{(L_L \to \nu_L)}(\mu)&=&
g_{L\nu}^2 D_Z(\mu)+\frac12D_W(\mu)\,,\nn
D_C^{(L_L \to E_L)}(\mu)&=&
g_{Le}^2 D_Z(\mu)+\frac12D_W(\mu)\,,\nn
D_C^{(E_R \to E_R)}(\mu)&=& g_{Re}^2 D_Z(\mu)\,,
\label{87}
\end{eqnarray}
where $g_{LU}=1/2-2/3 \sin^2\theta_W$, $g_{RU}=-2/3 \sin^2\theta_W$, etc.\ are the $Z$ charges of the fermions, and
\begin{eqnarray}
D_Z(\mu) &=& \frac{\alpha_Z}{4\pi}\Biggl(  2\log \frac{M_Z^2}{\mu^2} \log \frac{\bar n \cdot p}{\mu}- \frac12\log^2 \frac{M_Z^2}{\mu^2}\nn
&&\qquad - \frac32 \log \frac{M_Z^2}{\mu^2}-\frac{5\pi^2}{12}+ \frac94\Biggr)\,,\nn
D_W(\mu) &=& \frac{\alpha_W}{4\pi}\Biggl(2 \log \frac{M_W^2}{\mu^2} \log \frac{\bar n \cdot p}{\mu}- \frac12\log^2 \frac{M_W^2}{\mu^2}\nn
&&\qquad- \frac32 \log \frac{M_W^2}{\mu^2}-\frac{5\pi^2}{12}+ \frac94\Biggr)\,,
\end{eqnarray}
where $\alpha_W=\aem/\sin^2\theta_W$, $\alpha_Z = \aem/(\sin^2\theta_W \cos^2\theta_W)$.

The additional contributions for the $t,b$-quarks are given by
\begin{eqnarray}
F_{t_L}(\mu) &=& \left(\frac{\alpha_s}{4\pi} \frac{4}{3}  +\frac{\aem}{4\pi} \frac49\right)\nn
&&\times\left(\frac12 \log^2 \frac{m_t^2}{\mu^2}-\frac12 \log \frac{m_t^2}{\mu^2} + \frac{\pi^2}{12}+2\right)\nn
&&+\frac{\alpha_W}{4\pi} \frac12 f_F\left(\frac{m_t^2}{M_W^2},0\right) + \frac{\alpha_Z}{4\pi} g_{Lt}^2 f_F\left(\frac{m_t^2}{M_Z^2},\frac{m_t^2}{M_Z^2}\right)\nn
&&+\left(\delta \waver_{t_L}-\delta \waver_{u_L}\right)\,,
\nn
F_{t_R}(\mu) &=& \left(\frac{\alpha_s}{4\pi} \frac{4}{3}  +\frac{\aem}{4\pi} \frac49\right)\nn
&&\times\left(\frac12 \log^2 \frac{m_t^2}{\mu^2}-\frac12 \log \frac{m_t^2}{\mu^2} + \frac{\pi^2}{12}+2\right)\nn
&&+ \frac{\alpha_Z}{4\pi}  g_{Rt}^2 f_F\left(\frac{m_t^2}{M_Z^2},\frac{m_t^2}{M_Z^2}\right)+\left(\delta \waver_{t_R}-\delta \waver_{u_R}\right)\,,\nn
F_{b_L^\prime}(\mu) &=& \frac{\alpha_W}{4\pi} \frac12f_F\left(0,\frac{m_t^2}{M_W^2}\right)+\left(\delta \waver_{b_L}-\delta \waver_{d_L}\right)\,.
\end{eqnarray}
The $\alpha_s$ and $\aem$ terms are from the QCD and QED corrections due to the transition from SCET to bHQET fields. The functions $f_{F,S}$ are given in Appendix~B of Ref.~\cite{CKM}. $\left(\delta \waver_{t_L}-\delta \waver_{u_L}\right)$ is the difference in wavefunction corrections for the $t$ and a massless quark. The massless wavefunction contribution has already been included in $D_{W,Z}$.

\begin{figure}
\begin{center}
\includegraphics[width=3.75cm]{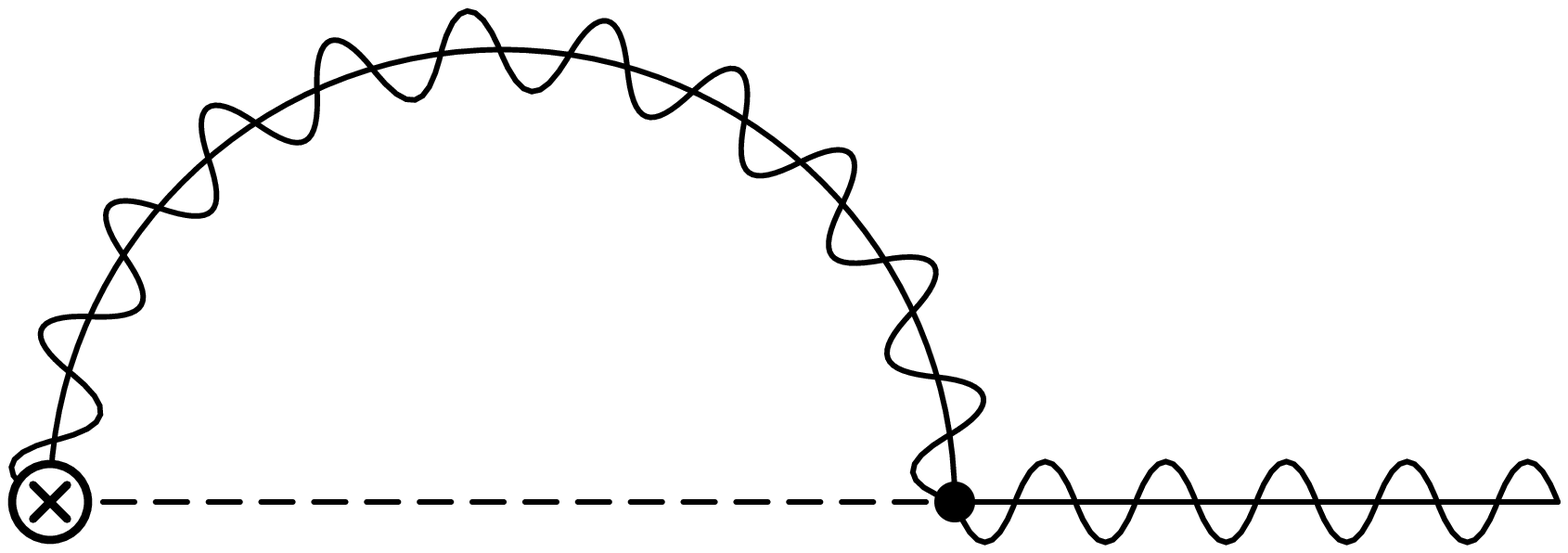} \qquad
\includegraphics[width=3.75cm]{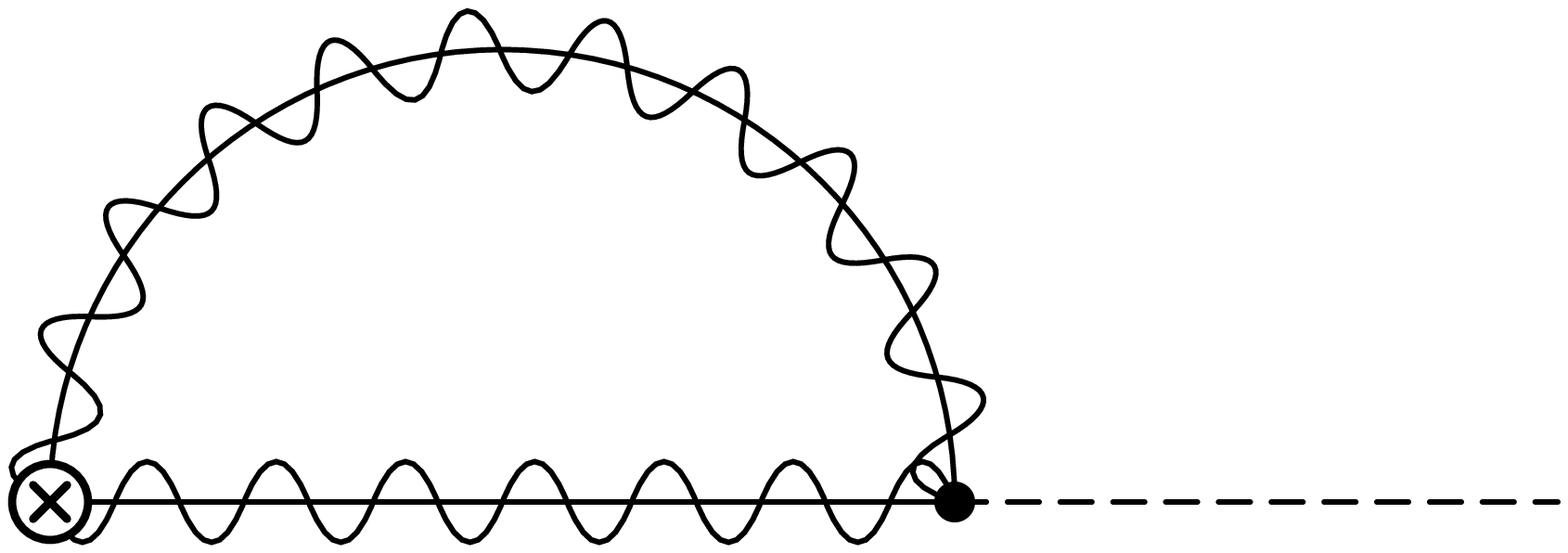} 
\end{center}
\caption{\label{fig:fd3} One loop collinear graphs which induce mixing between the gauge and Higgs sectors.}
\end{figure}

The $H$, $\varphi$ and gauge boson matching has mixing effects due to graphs such as those in Fig.~\ref{fig:fd3}. The graphs are of order $\vev{\phi}/(\bar n \cdot p)$ and are subleading in the SCET power counting.
%%%--FIGURE--------------------------------------------------------------------------------------
%%%-------------------------------------------------------------------------------------------------

The matching function for the Higgs doublet has some interesting features. The Higgs doublet is
\begin{eqnarray}
\phi &=& \frac{1}{\sqrt 2}\left[ \begin{array}{c}
\varphi^2 + i \varphi^1 \\
v + H - i \varphi_3 \end{array} \right]\,
\end{eqnarray}
with $\varphi^\pm = (\varphi^1 \mp i \varphi^2)/\sqrt2$. There are two neutral gauge bosons, the $Z$ and $\gamma$, but only one neutral unphysical Goldstone boson, the $\varphi^3$.
One could try a matching relation of the form
\begin{eqnarray}
\left[W^\dagger_{\text{EW}} \phi\right] & \rightarrow & \left[ \begin{array}{c} 
\exp D_C^{(\phi \to \varphi^+)}\ [W^\dagger_\gamma \varphi^+] \\
\frac{1}{\sqrt 2} \exp D_C^{(\phi \to H)}\   H-\frac{i}{\sqrt 2} \exp D_C^{(\phi \to \varphi^3)}\ \varphi^3 \\
\end{array}\right]\,,\nn
\label{varphimatch}
\end{eqnarray}
analogous to the fermionic case discussed above. A matching of this kind, which was used in Ref.~\cite{p1} for the $SU(2)$ theory, \emph{is not possible for the standard model}. The $\varphi^+$ propagator in the full theory has photon corrections
%%%----FIGURE--------------------------------------------------------------------------------------
\begin{figure}
\begin{center}
\includegraphics[width=4cm]{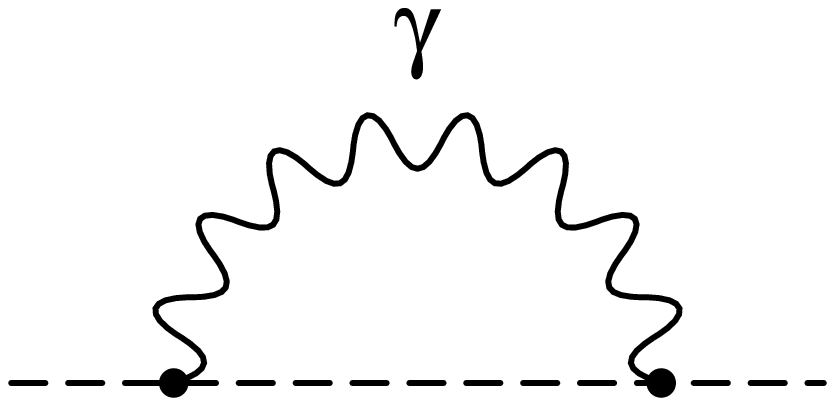} \\
\includegraphics[width=4cm]{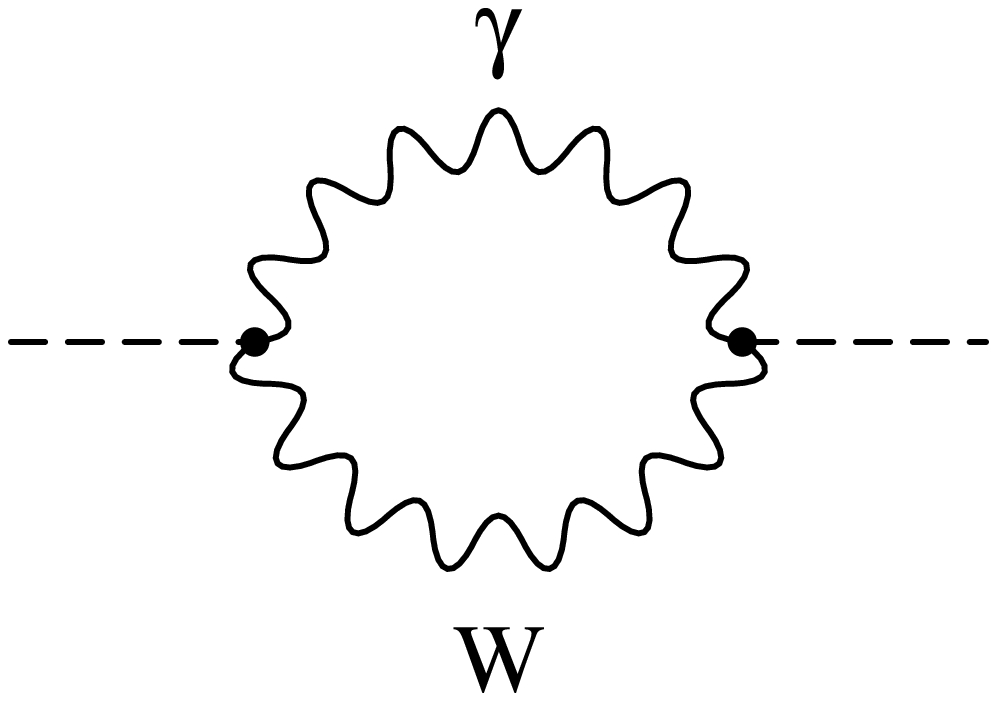} 
\end{center}
\caption{\label{fig:phiwave} Photon corrections to the $\varphi^+$ propagator.}
\end{figure}
%%
%------------------------------------------------------------------------------------------------------
shown in Fig.~\ref{fig:phiwave}. The graphs are infrared divergent, but the infrared divergence cancels between the two diagrams so that the $\varphi^+$ propagator is not infrared divergent in the electroweak theory. In the theory below $\mu_l$, the $W$ bosons have been integrated out, and the second diagram is absent, so that the $\varphi^+$ propagator is infrared divergent. Thus the infrared divergences do not match between the theories above and below $\mu_l$.

\begin{figure}
\begin{center}
\includegraphics[width=6cm]{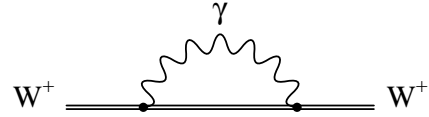} \\
\end{center}
\caption{\label{fig:w} Photon corrections to the bHQET $W^+$ propagator.}
\end{figure}

The resolution of this paradox is that $\varphi^+$ is not a physical field and is gauge dependent. At the scale $\mu_l$, the Higgs doublet matches, not to the Higgs $H$ and unphysical Goldstone bosons $\varphi^+$ and $\varphi^3$, but to $H$ and longitudinal gauge bosons $W_L$ and $Z_L$. $W_L$ is treated as a bHQET field, and the $W_L$ propagator has an infrared divergence from Fig.~\ref{fig:w}, so there is still an infrared
%%%----FIGURE--------------------------------------------------------------------------------------
%%
%------------------------------------------------------------------------------------------------------
divergence in the effective theory. However now, the amplitude that must be matched is for $W_L$, not $\varphi^+$, and is given by the amplitude for $\varphi^+$ multiplied by the equivalence theorem factor $\mathcal{E}$, which is the radiative correction factor in the equivalence theorem\cite{cornwall,vayonakis,leequiggthacker,chanowitz,gounaris,bagger,yaoyuan,bohmbook}. \emph{There is an infrared divergence in $\mathcal{E}$ that matches the infrared divergence in the effective theory.}  The standard model one-loop values for $\mathcal{E}_{W,Z}$ needed for  longitudinal $W$ and $Z$ production are given in Appendix~\ref{app:et}.

The matching Eq.~(\ref{varphimatch}) should instead be written as
\begin{eqnarray}
\left[W^\dagger_{\text{EW}} \phi\right] & \rightarrow & \left[ \begin{array}{c} 
\exp D_C^{(\phi \to W^+_L)}\ [W^\dagger_\gamma W_L] \\
\frac{1}{\sqrt 2} \exp D_C^{(\phi \to H)}\   H-\frac{i}{\sqrt 2} \exp D_C^{(\phi \to Z_L)}\ Z_L \\
\end{array}\right]\,.\nn
\label{phimatch}
\end{eqnarray}
The collinear functions are
\begin{eqnarray}
D_C^{(\phi \to W_L)} &=& 
\frac{\alpha_W}{4\pi} \frac14\left[F_W + f_S\left(1,\frac{M_H^2}{M_W^2}\right)\right]\nn
&&+\frac{\alpha_W}{4\pi} \frac14\left[F_W + f_S\left(1,\frac{M_Z^2}{M_W^2}\right)\right]\nn
&&+\frac{\alpha_Z}{4\pi} g_{\varphi^+}^2\left[F_Z + f_S\left(\frac{M_W^2}{M_Z^2},\frac{M_W^2}{M_Z^2}\right)\right]\nn
&&+\frac{\alpha_W}{4\pi}\sWsq\left[ \frac12 \log^2 \frac{M^2}{\mu^2}- \log \frac{M^2}{\mu^2}+\frac{\pi^2}{12}+2\right]\nn
&&+\frac12\delta \waver_{\varphi^+}+\log \mathcal{E}_W\,,\nn
D_C^{(\phi \to Z_L)} &=& 
\frac{\alpha_W}{4\pi} \frac12\left[F_W + f_S\left(\frac{M_Z^2}{M_W^2},1\right)\right]\nn
&&+\frac{\alpha_Z}{4\pi}\frac14\left[F_Z + f_S\left(1,\frac{M_H^2}{M_Z^2}\right)\right]\nn
&&+\frac12\delta  \waver_{\varphi^3}+\log \mathcal{E}_Z\,,\nn
D_C^{(\phi \to H)} &=& 
\frac{\alpha_W}{4\pi} \frac12\left[F_W + f_S\left(\frac{M_H^2}{M_W^2},1\right)\right]\nn
&&+\frac{\alpha_Z}{4\pi}\frac14\left[F_Z + f_S\left(\frac{M_H^2}{M_Z^2},1\right)\right]\nn
&&+\frac12\delta \waver_{H}\,,
\label{eq34}
\end{eqnarray}
where
\begin{eqnarray}
F_{W,Z} &=& 2\log \frac{M_{W,Z}^2}{\mu^2} \log \frac{\bar n \cdot p}{\mu}- \frac12\log^2 \frac{M_{W,Z} ^2}{\mu^2}\nn
&&\qquad -  \log \frac{M_{W,Z} ^2}{\mu^2}-\frac{5\pi^2}{12}+1\,,
\end{eqnarray}
and $\mathcal{E}_{W,Z}$ are the equivalence theorem factors for the $W$ and $Z$. The expression for $\mathcal{E}_W$ is the same as that for the $SU(2)$ theory given in Ref.~\cite{p1}. $\mathcal{E}_Z$ is given by a similar expression, see Ref.~\cite{bohmbook} for details. There are corrections to the equivalence theorem from $\gamma-Z$ mixing at two-loops, if one does not use background field gauge~\cite{denner}.

The gauge field collinear matching involves $\gamma-Z$ mixing. The collinear functions are defined by
\begin{eqnarray}
\left[W^\dagger g_\perp \right] & \to & \exp D_C^{(g \to g)} g_\perp\,,\nn
\left[W^\dagger W^\pm_\perp \right] & \to & \exp D_C^{(W \to W)} W^\pm_\perp\,,\nn
\left[W^\dagger W^3_\perp \right] & \to & \cW \exp D_C^{(W \to Z)} Z_\perp
+\sW \exp D_C^{(W \to\gamma)} A_\perp\,,\nn
\left[W^\dagger B_\perp \right] & \to & -\sW \exp D_C^{(B \to Z)} Z_\perp
+\cW \exp D_C^{(B \to\gamma)} A_\perp\,,\nn
\end{eqnarray}
($\sW=\sin\theta_W$, $\cW=\cos \theta_W$), so that all the collinear functions vanish at tree-level. The complications of $\gamma-Z$ mixing only enter the effective theory at the low-scale matching at $\mu_l$.

The gluon matching is
\begin{eqnarray}
D_C^{(g \to g)} &=& \frac{\alpha_s}{4\pi}\frac13 \log \frac{m_t^2}{\mu^2}\,.
\end{eqnarray}
There is a non-trivial gluon collinear matching from the top-quark vacuum polarization graph, since the top-quark is integrated out at the scale $\mu_l$ and is no longer a dynamical field. Processes involving external top quark can still be computed using bHQET fields for the top.

The other gauge-field collinear functions are
\begin{eqnarray}
 D_C^{(W \to W)} &=& \frac{\alpha_W}{4\pi}\cWsq\left[F_Z + f_S\left(\frac{M_W^2}{M_Z^2},\frac{M_W^2}{M_Z^2}\right)\right]\nn
&&+ \frac{\alpha_W}{4\pi}\cWsq \left[F_W + f_S\left(1,\frac{M_Z^2}{M_W^2}\right)\right]\nn
&&+ \frac{\alpha_W}{4\pi}\sWsq \left[F_W + f_S\left(1,0\right)\right]\nn
&&\hspace{-2cm}+\frac{\alpha_W}{4\pi}\sWsq\left[ \frac12 \log^2 \frac{M^2}{\mu^2}- \log \frac{M^2}{\mu^2}+\frac{\pi^2}{12}+2\right]\nn
&&+\frac12\delta  \waver_{W^+}\,,\nn
D_C^{(W \to Z)} &=& \frac{\alpha_W}{4\pi}2\left[F_W + f_S\left(\frac{M_Z^2}{M_W^2},1\right)\right]\nn
&&+\frac12\delta \waver_{Z}+\tan \theta_W \waver_{\gamma \to Z}\,,\nn
D_C^{(B \to Z)} &=&  \frac12 \delta \waver_{Z}-\cot \theta_W\waver_{\gamma \to Z}\,, \nn
D_C^{(W^3 \to \gamma)} &=& \frac{\alpha_W}{4\pi}2 \left[F_W + f_S\left(0,1\right)\right]\nn
&&+\frac12\delta \waver_{\gamma}+\cot \theta_W \waver_{Z \to \gamma}\,,\nn
D_C^{(B \to \gamma)} &=& \frac12\delta \waver_{\gamma}-\tan \theta_W \waver_{Z \to \gamma}\,.
\end{eqnarray}
The definitions of $\waver_{\gamma \to Z}$ and $\waver_{Z \to \gamma}$, which arise from $\gamma-Z$ mixing, are given in the appendix.

\subsection{Running below $\mu_l \sim M_Z$}\label{sec:runl}

The collinear anomalous dimensions for the running below $\mu_l \sim M_Z$ are listed below. The gauge coupling constants are the \msbar\ values in the theory with five dynamical quark flavors.

$(u,c)_{L,R},(\bar u,\bar c)_{L,R}$:
\begin{eqnarray}
\frac{1}{4\pi}\left(\frac43 \alpha_s+\frac 4 9 \aem  \right)\left(4 \log \frac{\bar n \cdot p}{\mu} - 3 \right)
\end{eqnarray}

$(d,s,b)_{L,R},(\bar d,\bar s, \bar b)_{L,R}$:
\begin{eqnarray}
\frac{1}{4\pi}\left(\frac43 \alpha_s+\frac 1 9 \aem  \right)\left(4 \log \frac{\bar n \cdot p}{\mu} - 3 \right)
\end{eqnarray}

$t$ treated as a bHQET field $h_v$:
\begin{eqnarray}
\frac{1}{4\pi}\left(\frac43 \alpha_s+\frac 4 9 \aem\right)\left(4 \log 2 \gamma - 2 \right)
\end{eqnarray}

$W^\pm_{T,L}$ treated as a bHQET field $h_v$:
\begin{eqnarray}
\frac{1}{4\pi}\left(\aem\right)\left(4 \log 2 \gamma - 2 \right)
\end{eqnarray}

$Z_{T,L}$ treated as a bHQET field $h_v$:
\begin{eqnarray}
0
\end{eqnarray}

$H$ treated as a bHQET field $h_v$:
\begin{eqnarray}
0
\end{eqnarray}

$(e,\mu,\tau)_{L,R},(\bar e,\bar \mu,\bar \tau)_{L,R}$:
\begin{eqnarray}
\frac{\aem}{4\pi}\left(4 \log \frac{\bar n \cdot p}{\mu} - 3 \right)
\end{eqnarray}

$(\nu_e,\nu_\mu,\nu_\tau)_{L,R},(\bar \nu_e,\bar \nu_\mu,\bar \nu_\tau)_{L,R}$:
\begin{eqnarray}
0
\end{eqnarray}

$g$:
\begin{eqnarray}
\frac{\alpha_s}{4\pi}\left(12 \log \frac{\bar n \cdot p}{\mu} -\frac{23}{3} \right)
\end{eqnarray}

$\gamma$:
\begin{eqnarray}
\frac{\aem}{4\pi}\left( \frac{80}{9} \right)\,.
\end{eqnarray}

\section{Soft Functions}\label{sec:soft}

The universal soft functions is
\begin{eqnarray}
U_S(n_i,n_j) &=& \log\frac{-n_i \cdot n_j-i0^+}2\,,
\label{eq:U}
\end{eqnarray}
in terms of which, the soft anomalous dimension and low-scale matching are
\begin{eqnarray}
\bm{\gamma}_S &=& \Gamma(\alpha(\mu))\left[- \sum_{\vev{ij}} \mathbf{T}_i \cdot 
\mathbf{T}_j \
U_S(n_i,n_j)\right]\,,\nn
\mathbf{D}_S &=& J(\alpha(\mu),\lM)\left[- \sum_{\vev{ij}}  \mathbf{T}_i \cdot 
\mathbf{T}_j \ U_S(n_i,n_j)\right]\,,
\label{152}
\end{eqnarray}
where, at one-loop,
\begin{eqnarray}
 \Gamma(\alpha(\mu)) &=& \frac{\alpha(\mu)}{4\pi} 4\,,\nn
J(\alpha(\mu),\lM) &=& \frac{\alpha(\mu)}{4\pi}2\log \frac{M^2}{\mu^2}\,.
\end{eqnarray}
The soft anomalous dimension is mass independent, but the soft matching depends on the gauge boson mass $M$. In the computations, we have used the three-loop value for $\Gamma$~\cite{moch:ff}, and the results of Refs.~\cite{aybat1,aybat2}.

The soft function has a simple form when written using the color-operator notation~\cite{catani}. For practical calculations, one needs to write the soft function as a matrix in the space of gauge invariant operators. In this section, we give the explicit matrices needed for some scattering processes. The QCD parts of these matrices have been obtained previously~\cite{kidonakis}. The electroweak part is considerably more involved, because the $SU(2) \times U(1)$ symmetry is broken, and this enters into the low-scale soft function $\mathbf{D}_S$.

For the standard model, one has to use Eq.~(\ref{su2u1},\ref{su2u1a}) to obtain the soft anomalous dimension and low-scale matching. For a given process, the $SU(3)$, $SU(2)$ and $U(1)$ matrices are defined by
\begin{eqnarray}
\softm_3 &=& - \sum_{\vev{ij}} \mathbf{T}_i \cdot \mathbf{T}_j\ U_S(n_i,n_j)\,,\nn
\softm_2 &=& - \sum_{\vev{ij}} \mathbf{t}_i \cdot \mathbf{t}_j\ U_S(n_i,n_j)\,,\nn
\softm_1 &=& - \sum_{\vev{ij}} Y_i Y_j\ U_S(n_i,n_j)\,,
\end{eqnarray}
in terms of which, the soft anomalous dimension is
\begin{eqnarray}
\bm{\gamma}_S &=& \frac{\alpha_s}{\pi} \softm_3+\frac{\alpha_2}{\pi}\softm_2
+\frac{\alpha_1}{\pi}\softm_1\,.
\label{softad}
\end{eqnarray}

For the low-scale matching, one has to use Eq.~(\ref{su2u1},\ref{su2u1a}) with $M \to M_W$ and $M \to M_Z$ for the $W$ and $Z$ terms. This gives
\begin{eqnarray}
\mathbf{D}_S &=& \frac{\alpha_W(\mu)}{4\pi}2\log \frac{M_W^2}{\mu^2}\left[- \sum_{\vev{ij}}  \left(\mathbf{t}_i \cdot \mathbf{t}_j-t_{3i}t_{3j}\right) \ U_S(n_i,n_j)\right]\nn
&&+ \frac{\alpha_Z(\mu)}{4\pi}2\log \frac{M_Z^2}{\mu^2}\left[- \sum_{\vev{ij}}  t_{Zi}t_{Zj} \ U_S(n_i,n_j)\right]\nn
&=& \frac{\alpha_W(\mu)}{4\pi}2\log \frac{M_W^2}{\mu^2}\left[\softm_2 + \sum_{\vev{ij}}  t_{3i}t_{3j} \ U_S(n_i,n_j)\right]\nn
&&+ \frac{\alpha_Z(\mu)}{4\pi}2\log \frac{M_Z^2}{\mu^2}\left[- \sum_{\vev{ij}}  t_{Zi}t_{Zj} \ U_S(n_i,n_j)\right]\,.
\label{152.a}
\end{eqnarray}

The soft function has a  universal form when written in the operator form Eq.~(\ref{152}). For numerical computations, it is more convenient to choose a basis of gauge invariant operators, and write the soft-anomalous dimension and matching as a matrix in the chosen basis. The soft factor $\sum_{\vev{ij}} \mathbf{T}_i \cdot \mathbf{T}_j\ U_S(n_i,n_j)$ was computed for some simple cases in Ref.~\cite{p1} for an $SU(N)$ gauge theory. Certain soft matrices occur in several different computations. These reference matrices are for SU(3):
\begin{eqnarray}
\softm^{(3)}&=&-\frac{8}{3}i\pi  \openone+ \left[ \begin{array}{cc} 
\frac73 T +\frac23 U & 2(T-U) \\
\frac{4}{9} (T-U)
 & 0
 \end{array} \right]\,,\nn
 \softm^{(3)\, \prime} &=& -\frac43 i \pi\,,\nn
 \softm^{(3,g)}&=& -\frac{13}{3}i\pi  \openone+\left[ \begin{array}{ccc}
0 & 0 & U-T \\
0 & \frac32(T+U) & \frac32 (U-T)\\ 
2  (U-T)  & \frac56  (U-T) & \frac32(T+U)\\ 
 \end{array} \right]\nn
 \label{m3}
\end{eqnarray}
for $SU(2)$:
\begin{eqnarray}
\softm^{(2)}
&=&-\frac32i\pi \openone+ \left[ \begin{array}{cc} 
 (T+U)  & 2(T-U) \\
\frac38 (T-U)
 & 0
 \end{array} \right]\,,\nn
 \softm^{(2)\, \prime} &=& -\frac34 i \pi\,,\nn
\softm^{(2,g)}&=&-\frac{11}{4}i\pi \openone+\left[ \begin{array}{ccc}
0  & U-T \\
2  (U-T) & (T+U)\\ 
 \end{array} \right]
 \label{m2}
\end{eqnarray}
and for $U(1)$:
\begin{eqnarray}
\softm^{(1)}(q_1,q_2,q_3,q_4)
&=& 
 -i  \frac{\pi} 2\left(q_1^2+q_2^2+q_3^2+q_4^2 \right)\nn
 &&+ \left(q_1 q_4 +q_2 q_3\right) T -\left( q_1 q_3+q_2 q_4 \right) U \,,\nn
\softm^{(1)}(q_f,q_i) &=& 
 -i  \pi \left(q_i^2+q_f^2 \right)+ 2 q_i q_f \left(T-U \right)\,.\nn
 \label{m1}
\end{eqnarray}

For scattering kinematics, $s>0$, $t<0$, and $u<0$, and the variables $T,U$ are defined by~\cite{kidonakis}
\begin{eqnarray}
T &=& \log \frac{-t}{s} + i \pi\,,\nn
U &=& \log \frac{-u}{s} + i \pi\,.
\end{eqnarray}

\section{Soft Functions for Fermion Scattering}\label{sec:fermions}

The soft anomalous dimension and low-scale matching matrices will now be computed for some scattering processes.
In these examples, the anomalous dimension and matching depend on matrices $\softm_{3,2,1}$, $R^{(0)}$, and $\softd_{W,Z}$.
The equations for the anomalous dimension and matching have the same form in each case; the matrices take on different values depending on the process.

\subsection{Two doublets}

Consider first the case of fermion scattering, $Q \bar Q \to Q \bar Q$, where all particles are electroweak doublet quarks. At the high-scale, one matches onto four-quark SCET operators
\begin{eqnarray}
&& C_{11}\ \bar Q^{(c)}_4 t^a T^A \gamma^\mu P_L Q^{(c)}_3\  \bar Q^{(u)}_2 t^a T^A\gamma^\mu P_L Q^{(u)}_1\nn
&+&C_{21}\ \bar Q^{(c)}_4  \gamma^\mu T^A P_L Q^{(c)}_3 \ \bar Q^{(u)}_2 \gamma^\mu T^A P_L Q^{(u)}_1\nn
&+& C_{12}\ \bar Q^{(c)}_4 t^a \gamma^\mu P_L Q^{(c)}_3 \ \bar Q^{(u)}_2 t^a \gamma^\mu P_L Q^{(u)}_1\nn
&+&C_{22}\ \bar Q^{(c)}_4  \gamma^\mu P_L Q^{(c)}_3\  \bar Q^{(u)}_2 \gamma^\mu P_L Q^{(u)}_1\,,
\label{85}
\end{eqnarray}
where the first index is $1$ for $t^a \otimes t^a$ and $2$ for $\mathbf{1} \otimes \mathbf{1}$ in $SU(2)$,
and the second index is $1$ for $T^a \otimes T^a$ and $2$ for $\mathbf{1} \otimes \mathbf{1}$ in $SU(3)$. Eq.~(\ref{85}) is written in schematic form to emphasize the gauge structure of the operator. The actual operator in SCET should be written with $Q \to W^\dagger \xi_{n,p}^{(Q)}$, etc. The subscripts $1-4$ on the fields are a reminder that the SCET fields have momentum labels $p_1-p_4$. We have chosen to label the two fields by $u$ and $c$ to make it easy to discuss related processes such as Drell-Yan by replacing some quark fields by lepton fields. The one-loop values for $C_{ij}$ at the high scale are given in Ref.~\cite{CKM}.

The group theory sums needed for the soft anomalous dimension matrix are
\begin{eqnarray}
\softm_3&=& -\sum_{\vev{ij}} \mathbf{T}_i \cdot 
\mathbf{T}_j\ U_S(n_i,n_j)
= \openone \otimes \softm^{(3)}\,,\nn
\softm_2&=& -\sum_{\vev{ij}} \mathbf{t}_i \cdot 
\mathbf{t}_j\ U_S(n_i,n_j)= \softm^{(2)} \otimes  \openone\,,\nn
 \softm_1&=&- \sum_{\vev{ij}} Y_i \, Y_j U_S(n_i,n_j) \nn
 &=&  \softm^{(1)}\left(Y(Q^{(c)}),Y(Q^{(u)})\right)\openone \otimes \openone\,,
 \label{158a}
\end{eqnarray}
in terms of the reference matrices $\softm^{(3,2,1)}$ given in Eqs.~(\ref{m3},\ref{m2},\ref{m1}). For quark doublets,  $Y(Q^{(c)})=Y(Q^{(u)})=1/6$.

The soft anomalous dimension is given by Eq.~(\ref{softad}) using Eq.~(\ref{158a}) for $\softm_{3,2,1}$.
At the low scale $\mu_l \sim m_Z$, the operators Eq.~(\ref{85})
match onto a linear combination of
\begin{eqnarray}
\widehat{\mathcal{O}}_{11} &=& [\bar{c}_{L4} T^A  \gamma_\mu c_{L3}] [\bar{u}_{L2} T^A \gamma^\mu u_{L1}]  \nn
\widehat{\mathcal{O}}_{21} &=&  [\bar{c}_{L4}T^A   \gamma_\mu c_{L3}][\bar{d}_{L2}^\prime T^A \gamma^\mu d_{L1}^\prime] \nn
\widehat{\mathcal{O}}_{31} &=& [\bar{s}_{L4}^\prime T^A \gamma_\mu s_{L3}^\prime] [\bar{u}_{L2}  T^A\gamma^\mu  u_{L1}]  \nn
\widehat{\mathcal{O}}_{41} &=& [\bar{s}_{L4}^\prime T^A \gamma_\mu s_{L3}^\prime] [\bar{d}_{L2}^\prime  T^A\gamma^\mu  d_{L1}^\prime] \nn
\widehat{\mathcal{O}}_{51} &=&   [\bar{s}_{L4}^\prime T^A \gamma_\mu  c_{L3}][\bar{u}_{L2}T^A\gamma^\mu   d_{L1}^\prime]\nn
\widehat{\mathcal{O}}_{61} &=&  [\bar{c}_{L4}  T^A \gamma_\mu s_{L3}^\prime][\bar{d}_{L2}^\prime  T^A\gamma^\mu  u_{L1}] \nn
\widehat{\mathcal{O}}_{12} &=& [\bar{c}_{L4}  \gamma_\mu c_{L3}] [\bar{u}_{L2}  \gamma^\mu u_{L1}]  \nn
\widehat{\mathcal{O}}_{22} &=&  [\bar{c}_{L4}   \gamma_\mu c_{L3}][\bar{d}_{L2}^\prime  \gamma^\mu d_{L1}^\prime] \nn
\widehat{\mathcal{O}}_{32} &=& [\bar{s}_{L4}^\prime   \gamma_\mu s_{L3}^\prime] [\bar{u}_{L2} \gamma^\mu  u_{L1}]  \nn
\widehat{\mathcal{O}}_{42} &=& [\bar{s}_{L4}^\prime   \gamma_\mu s_{L3}^\prime] [\bar{d}_{L2}^\prime \gamma^\mu  d_{L1}^\prime] \nn
\widehat{\mathcal{O}}_{52} &=&   [\bar{s}_{L4}^\prime   \gamma_\mu c_{L3}][\bar{u}_{L2} \gamma^\mu  d_{L1}^\prime]\nn
\widehat{\mathcal{O}}_{62} &=&  [\bar{c}_{L4}   \gamma_\mu s_{L3}^\prime][\bar{d}_{L2}^\prime\gamma^\mu   u_{L1}] 
\label{81}
\end{eqnarray}
with coefficients $\widehat C_{ij}$. The matching matrix is
\begin{eqnarray}
\widehat C_{ia} &=& R_{ij} C_{ja}\,,
\label{87a}
\end{eqnarray}
or equivalently,
\begin{eqnarray}
\widehat C &=& \left( R \otimes \openone \right) C\,,
\label{87b}
\end{eqnarray}
since the electroweak matching does not change the color structure of the operators.
At tree-level
$R$ is
\begin{eqnarray}
R^{(0)} &=& \left[\begin{array}{rc} \frac14 &1 \\[5pt]
-\frac14 & 1\\[5pt]
 -\frac14 & 1 \\[5pt]
 \frac14 & 1 \\[5pt]
\frac12 & 0  \\[5pt]
\frac12 & 0  
\end{array}\right]\,.
\label{88}
\end{eqnarray}
Once again, we see the additional complication in the standard model due to the $U(1)$ sector. In the pure $SU(2)$ theory, the matching was $SU(2)$ invariant; here the operators have to be broken apart into individual fields of definite charge.

The one-loop soft matching due to $W$ and $Z$ exchange is computed using Eq.~(\ref{152.a}),
\begin{eqnarray}
R_{S,W}^{(1)} &=& \frac{\alpha_W}{4\pi} 2\log \frac{M_W^2}{\mu^2}\left[\softm_2+ \sum_{\vev{ij}} t_{3i}t_{3j} U_S(n_i,n_j)\right]\,,\nn
R_{S,Z}^{(1)} &=&\frac{\alpha_Z}{4\pi} 2\log \frac{M_Z^2}{\mu^2}
 \left[-  \sum_{\vev{ij}} t_{Zi}t_{Zj}U_S(n_i,n_j)\right]\,,
 \label{168}
\end{eqnarray}
and the total soft matching at one-loop is
\begin{eqnarray}
R &=&R^{(0)}+R_{S,W}^{(1)}+R_{S,Z}^{(1)}\,.
\end{eqnarray}

To evaluate $R_{S,W}^{(1)},R_{S,Z}^{(1)}$, we need to evaluate the group theory factors in Eq.~(\ref{168}). The $\softm_2$ term acting on Eq.~(\ref{81}) is a group-invariant Casimir operator, and can be thought of as $\softm_2$ acting on the original basis Eq.~(\ref{85}) before $SU(2) \times U(1)$ breaking, and so acting on the low-energy basis Eq.~(\ref{81}) is equal to $R^{(0)} \softm_2$, where $\softm_2$ is the matrix in Eq.~(\ref{158a}). The $t_{3i} t_{3j}$ and $t_{Zi}t_{Zj}$ terms are diagonal in the basis Eq.~(\ref{81}), and we define them to be $\softd_W$ and $\softd_Z$, respectively, so that the soft-matching matrices are
\begin{eqnarray}
R_{S,W}^{(1)} &=&\frac{\alpha_W}{4\pi} 2\log \frac{M_W^2}{\mu^2} \left[R^{(0)}\softm_2+\softd_W R^{(0)}\right]\,,\nn
R_{S,Z}^{(1)} &=& \frac{\alpha_Z}{4\pi} 2\log \frac{M_Z^2}{\mu^2} \left[\softd_Z R^{(0)}\right]\,.
 \label{167a}
\end{eqnarray}
This equation is valid for all the scattering processes we will consider. The $W$ matching has a $\softm_2$ term, which is the same matrix that enters the soft anomalous dimension, and the $W$ and $Z$ matchings have extra diagonal matrices $\softd_{W,Z}$ that depend on the process.

For the doublet scattering case, Eq.~(\ref{81}), $\softd_{W,Z}$ are:
\begin{eqnarray}
\softd_W &=& \text{diag}(w_1, -w_1 , -w_1 , w_1 , w_2 ,w_2)+\frac12 i \pi \openone\,, \nn
\softd_Z &=& \text{diag}(z_1,z_2,z_3,z_4,z_5,z_5)\,, \nn
w_1 &=& -\frac12(T-U)\,, \nn
w_2 &=&- \frac12(T+U)\,, \nn
z_1 &=&  2g_{Lc}g_{Lu}\left(T-U\right)-i\pi \left( g_{Lc}^2+g_{Lu}^2\right)\,, \nn
z_2 &=&  2g_{Lc}g_{Ld}\left(T-U\right)-i\pi \left( g_{Lc}^2+g_{Ld}^2\right)\,, \nn
z_3 &=&  2g_{Ls}g_{Lu}\left(T-U\right)-i\pi \left( g_{Ls}^2+g_{Lu}^2\right)\,, \nn
z_4 &=&  2g_{Ls}g_{Ld}\left(T-U\right)-i\pi \left( g_{Ls}^2+g_{Ld}^2\right)\,, \nn
z_5 &=& \left(g_{Lu}g_{Lc}+g_{Ld}g_{Ls}\right)T-\left( g_{Lu}g_{Ls}+g_{Ld}g_{Lc}\right) U\nn
&&-\frac{i}{2}\pi\left( g_{Lc}^2+g_{Lu}^2+ g_{Ls}^2+g_{Ld}^2\right)\,.
\label{171}
\end{eqnarray}

The results Eq.~(\ref{167a},\ref{171}) hold for all cases where both fermions are doublets. For example, if the final quark doublet is replaced by a lepton doublet, one gets four-fermion operators for the Drell-Yan process $q \bar q \to \mu^+ \mu^-$.   The four-quark operators only have the tensor structure $\mathbf{1} \otimes \mathbf{1}$ in color space and the anomalous dimension is Eq.~(\ref{softad}) with $\softm^{(3)} \to \softm^{(3)\, \prime}$ and $Y(Q^{(c)}) \to Y(L^{(\mu)})=-1/2$ in Eq.~(\ref{158a}). The unit matrix in color space is now a $1 \times 1$ matrix instead of a $2 \times 2$ matrix. The low-scale matching is obtained from Eq.~(\ref{171}) with the obvious replacement $g_{Lc} \to g_{L\nu}$, $g_{Ls} \to g_{Le}$. A similar result holds if the initial doublet is a lepton doublet and the final is a quark doublet, or if both are lepton doublets (in which case, $\softm_3 \to 0$).

\subsection{One doublet and one singlet}

The second case is where one fermion is a doublet and the other is a singlet. As an example, consider 
\begin{eqnarray}
&& C_{1} \bar Q^{(c)}_4  \gamma^\mu T^A P_L Q^{(c)}_3  \bar  u_2  \gamma^\mu  T^A P_R u_1\nn
&+& C_{2} \bar Q^{(c)}_4  \gamma^\mu P_L Q^{(c)}_3  \bar u_2 \gamma^\mu P_R u_1
\end{eqnarray}
The group theory sums needed for the soft anomalous dimension matrix are
\begin{eqnarray}
\softm_3&=& -\sum_{\vev{ij}} \mathbf{T}_i \cdot 
\mathbf{T}_j\ U_S(n_i,n_j)= \softm^{(3)}\,, \nn
\softm_2&=& -\sum_{\vev{ij}} \mathbf{t}_i \cdot 
\mathbf{t}_j\ U_S(n_i,n_j)
=  \softm^{(2)\, \prime}\,, \nn
\softm_1&=&- \sum_{\vev{ij}} Y_i \, Y_j U_S(n_i,n_j)
=  \softm^{(1)}\left(Y(Q^{(c)}),Y(u_R)\right)\,.\nn
\label{158b}
\end{eqnarray}

At the low scale $\mu_l \sim m_Z$, the operators Eq.~(\ref{85})
match onto a linear combination of
\begin{eqnarray}
\widehat{\mathcal{O}}_{11} &=& [\bar{c}_{L4}  T^A \gamma_\mu c_{L3}] [\bar{u}_{R2} T^A \gamma^\mu u_{R1}]  \nn
\widehat{\mathcal{O}}_{21} &=& [\bar{s}_{L4}^\prime T^A \gamma_\mu s_{L3}^\prime] [\bar{u}_{R2} T^A \gamma^\mu  u_{R1}]  \nn
\widehat{\mathcal{O}}_{12} &=& [\bar{c}_{L4}   \gamma_\mu c_{L3}] [\bar{u}_{R2} \gamma^\mu u_{R1}]  \nn
\widehat{\mathcal{O}}_{22} &=& [\bar{s}_{L4}^\prime  \gamma_\mu s_{L3}^\prime] [\bar{u}_{R2} \gamma^\mu  u_{R1}]  \,.
\label{97}
\end{eqnarray}
The matching matrix is
\begin{eqnarray}
\widehat C_{ia}&=& R_{i} \, C_a \Rightarrow \widehat C = \left(R \otimes \openone\right)C\,,
\label{98}
\end{eqnarray}
since the matching leaves the color structure unchanged.
At tree-level
$R$ is
\begin{eqnarray}
R^{(0)} &=& \left[\begin{array}{cc} 
 1\\
1 \\
\end{array}\right]\,.
\label{99}
\end{eqnarray}
At one-loop, the soft matching matrices due to $W$ and $Z$ exchange are
\begin{eqnarray}
\softd_W &=& \text{diag}(w_1, w_1)\,, \nn
\softd_Z &=& \text{diag}(z_1,z_2)\,, \nn
w_1 &=& \frac14i\pi\,, \nn
z_1 &=&  2g_{Lc}g_{Ru}\left(T-U\right)-i\pi \left( g_{Lc}^2+g_{Ru}^2\right)\,, \nn
z_2 &=&  2g_{Ls}g_{Ru}\left(T-U\right)-i\pi \left( g_{Ls}^2+g_{Ru}^2\right)\,,\nn
\label{171a}
\end{eqnarray}
and the soft matching is given by Eq.~(\ref{167a}).

Equations~(\ref{98}), (\ref{99}), (\ref{171a}) apply to all cases where one fermion is a weak doublet, and the other is a weak singlet, with the obvious replacement of the $Z$ charges for lepton doublets.

\subsection{Two singlets}

The last case is if both fermions are weak singlets, for example the operators
\begin{eqnarray}
&& C_1\ \bar c_4  \gamma^\mu T^A P_R c_3 \, u_2  \gamma^\mu T^A  P_R u_1\nn
&+&C_2\ \bar c_4  \gamma^\mu P_R c_3 \, u_2  \gamma^\mu P_R u_1
\end{eqnarray}
which match to
\begin{eqnarray}
\widehat{\mathcal{O}}_1 &=& [\bar{c}_{R4}  \gamma_\mu T^A c_{R3}] [\bar{u}_{R2}  \gamma^\mu T^A u_{R1}]\nn
\widehat{\mathcal{O}}_2 &=& [\bar{c}_{R4}  \gamma_\mu c_{R3}] [\bar{u}_{R2}  \gamma^\mu u_{R1}]  \,.
\label{106}
\end{eqnarray}
The group theory sums needed for the soft anomalous dimension matrix are
\begin{eqnarray}
\softm_3&=& -\sum_{\vev{ij}} \mathbf{T}_i \cdot 
\mathbf{T}_j\ U_S(n_i,n_j)= \softm^{(3)}\,, \nn
\softm_2&=& -\sum_{\vev{ij}} \mathbf{t}_i \cdot 
\mathbf{t}_j\ U_S(n_i,n_j)
=  0\,, \nn
\softm_1&=&- \sum_{\vev{ij}} Y_i \, Y_j U_S(n_i,n_j)
=  \softm^{(1)}\left(Y(c_R),Y(u_R)\right)\,, \nn
\label{158c}
\end{eqnarray}
which are used in Eq.~(\ref{softad}) to obtain the soft anomalous dimension.

The one-loop matching condition is 
\begin{eqnarray}
\widehat C_a = R C_a \Rightarrow \widehat C = \left(R \otimes \openone\right)C\,,
\end{eqnarray}
 with $R^{(0)}=1$ at tree-level, and the soft matching contribution is
\begin{eqnarray}
R_{S,W}^{(1)} &=& 0\,, \nn
R_{S,Z}^{(1)}  &=& \frac{\alpha_Z}{4\pi}
2 \log\frac{M_Z^2}{\mu^2} \times \nn
&&\Bigl[\left(T-U\right) 2g_{Rc}g_{Ru}-i\pi \left(g_{Rc}^2+g_{Ru}^2 \right)\Bigr]\,.\nn
\label{107}
\end{eqnarray}

One can similarly obtain the results for right-handed leptons by replacing the quark $Z$-charges by the corresponding lepton charges.

\section{Soft Functions for Electroweak Gauge Boson Pair Production}\label{sec:Wprod}

\subsection{Doublets}

\begin{figure}
\begin{center}
\includegraphics[width=6cm]{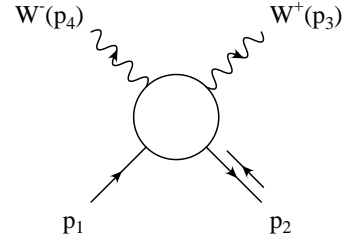}
\end{center}
\caption{\label{fig:W} Pair production $q(p_1) + \bar{q}(p_2) \to  W^+(p_3)+W^-(p_4)$. Time runs vertically. }
\end{figure}

The kinematics for the electroweak gauge boson pair-production is shown in Fig.~\ref{fig:W}.
%%%----FIGURE--------------------------------------------------------------------------------------
%%%------------------------------------------------------------------------------------------------------

We first start with gauge boson production by left-handed quarks, which are electroweak doublets, and interact with the $W$ and $B$ gauge bosons of the $SU(2)$ and $U(1)$ interactions.
The operator basis is
\begin{eqnarray}
O_1 &=& \bar Q^{(u)}_2 Q^{(u)}_1 W^a_4W^a_3\nn
O_2 &=& \bar Q^{(u)}_2 t^c Q^{(u)}_1  i \epsilon^{abc} W^a_4 W^b_3 \nn
O_3 &=& \bar Q^{(u)}_2 t^a Q^{(u)}_1  B_4 W^a_3\nn
O_4 &=& \bar Q^{(u)}_2 t^a Q^{(u)}_1 W^a_4 B_3 \nn
O_5 &=& \bar Q^{(u)}_2  Q^{(u)}_1 B_4 B_3 
\label{185.b}
\end{eqnarray}
where only the gauge structure has been shown. The operators at the high scale are best written in terms of the $SU(2)$ and $U(1)$ gauge fields $W$ and $B$, rather than the mass eigenstate fields $Z$ and $\gamma$. Note that $\epsilon^{abc}W^a_3 W^b_4\not=0$ since the two $W$ fields have momentum labels $p_3$ and $p_4$ which are different.
In this basis
\begin{eqnarray}
\softm_3 &=& -\sum_{\vev{ij}} \mathbf{T}_i \cdot \mathbf{T}_j\ U_S(n_i,n_j) =\softm^{(3)\,\prime} \otimes \openone \,, \nn
\softm_2 &=& -\sum_{\vev{ij}} \mathbf{t}_i \cdot \mathbf{t}_j\ U_S(n_i,n_j) = 
\text{diag}(\softm_{2a},\softm_{2b},\softm_{2b},\softm_{2c})\,, \nn
\softm_{2a} &=&  \left(-i\pi \frac{11}4\right)\openone+\left[ \begin{array}{cc}
0 & U-T\\
2(U-T) & T+U \\
\end{array}\right]\,, \nn
\softm_{2b}  &=&  -\frac74 i\pi+U+T\,, \nn
\softm_{2c} &=&  -\frac34 i\pi  \,, \nn
\softm_1 &=& -\sum_{\vev{ij}} Y_i Y_j\ U_S(n_i,n_j) = \softm^{(1)}\left(0,Y(Q^{(u)})\right) \openone\,,
\label{eq186}
\end{eqnarray}
and the soft anomalous dimension between the scales $Q$ and $\mu_l \sim M_Z$ is given by Eq.~(\ref{softad}), where the matrices $\softm_{(3,2,1)}$ are given by Eqs.~(\ref{eq186}).

At the low-scale $\mu_l \sim M_Z$, the operators Eq.~(\ref{185.b}) match onto
\begin{eqnarray}
\widehat O_1 &=& \bar u_{L2} u_{L1} W^+_4W^-_3 \nn
\widehat O_2 &=& \bar u_{L2} u_{L1} W^-_4 W^+_3\nn
\widehat O_3 &=& \bar u_{L2} u_{L1} Z_4 Z_3  \nn
\widehat O_{4} &=& \bar u_{L2} u_{L1} A_4 Z_3 \nn
\widehat O_{5} &=& \bar u_{L2} u_{L1} Z_4 A_3\nn
\widehat O_{6} &=& \bar u_{L2} u_{L1}  A_4 A_3\nn
\widehat O_7 &=& \bar d_{L2} d_{L1}W^+_4  W^-_3 \nn
\widehat O_8 &=& \bar d_{L2} d_{L1}  W^-_4 W^+_3 \nn
\widehat O_9 &=& \bar d_{L2} d_{L1}Z_4  Z_3 \nn
\widehat O_{10} &=& \bar d_{L2} d_{L1} A_4 Z_3 \nn
\widehat O_{11} &=& \bar d_{L2} d_{L1} Z_4 A_3\nn
\widehat O_{12} &=& \bar d_{L2} d_{L1}  A_4 A_3\nn
\widehat O_{13} &=& \bar u_{L2} d_{L1}W^+_4 Z_3  \nn
\widehat O_{14} &=& \bar u_{L2} d_{L1} W^+_4 A_3 \nn
\widehat O_{15} &=& \bar u_{L2} d_{L1} Z_4  W^+_3 \nn
\widehat O_{16} &=& \bar u_{L2} d_{L1}  A_4 W^+_3\nn
\widehat O_{17} &=& \bar d_{L2} u_{L1}Z_4  W^-_3 \nn
\widehat O_{18} &=& \bar d_{L2} u_{L1}A_4  W^-_3 \nn
\widehat O_{19} &=& \bar d_{L2} u_{L1}W^-_4  Z_3 \nn
\widehat O_{20} &=& \bar d_{L2} u_{L1} W^-_4 A_3 \,.
\label{188.a}
\end{eqnarray}
The subscripts $3,4$ represent outgoing label momenta $p_3$ and $p_4$, and the gauge indices are to be treated as those on a quantum field, i.e.\ they represent the charge on the annihilation operator. These operators are to be treated in the same manner as terms in a Lagrangian. Thus $e \bar e \to W^+(k_1) W^-(k_2)$ is given by $\widehat C_4$ with $p_4=k_1$ and $p_3=k_2$, plus $\widehat C_5$ with $p_4=k_1$ and $p_3=k_2$.

The tree-level matching is
\begin{eqnarray}
 \widehat C_i &=&  \left(R^{(0)}\right)_{ij} C_j\,, \nn
 R^{(0)} &=&
\left[\begin{array}{ccccc} 
1 & \frac12  & 0 & 0 & 0\\ %1
1 & -\frac12 & 0 & 0 & 0\\ %2
\cWsq & 0 & -\frac12 \sW \cW &  -\frac12 \sW \cW & \sWsq \\ %3
\sW \cW & 0 & \frac12 \cWsq & -\frac12 \sWsq & - \sW \cW \\ %11
\sW \cW & 0 & -\frac12 \sWsq & \frac12 \cWsq & - \sW \cW \\ %12
\sWsq & 0 & \frac12 \sW \cW & \frac12 \sW \cW & \cWsq\\ %19
1 & -\frac12 & 0 & 0 & 0\\ %4
1 & \frac12 & 0 & 0 & 0\\ %5
\cWsq & 0 & \frac12 \sW \cW &  \frac12 \sW \cW & \sWsq \\ %3
\sW \cW & 0 & -\frac12 \cWsq & \frac12 \sWsq & - \sW \cW \\ %11
\sW \cW & 0 & \frac12 \sWsq & -\frac12 \cWsq & - \sW \cW \\ %12
\sWsq & 0 &-\frac12 \sW \cW & -\frac12 \sW \cW & \cWsq\\ %19
0 & -\frac{1}{\sqrt 2}\cW & 0 & -\frac{1}{\sqrt 2}\sW & 0\\ %7
0 & -\frac{1}{\sqrt 2}\sW & 0 & \frac1{\sqrt 2}\cW & 0\\ %15
0 & \frac{1}{\sqrt 2}\cW &  -\frac{1}{\sqrt 2}\sW & 0 & 0\\ %7
0 & \frac{1}{\sqrt 2}\sW & \frac1{\sqrt 2}\cW & 0 & 0 \\ %15
0 & -\frac{1}{\sqrt 2}\cW &  -\frac{1}{\sqrt 2}\sW & 0 & 0\\ %7
0 & -\frac{1}{\sqrt 2}\sW & \frac1{\sqrt 2}\cW & 0 & 0 \\ %15
0 & \frac{1}{\sqrt 2}\cW & 0 & -\frac{1}{\sqrt 2}\sW & 0\\ %7
0 & \frac{1}{\sqrt 2}\sW & 0 & \frac1{\sqrt 2}\cW & 0\\ %15
\end{array}\right]\,.\nn
\end{eqnarray}

The one-loop soft matching is given by Eq.~(\ref{167a}) where
\begin{eqnarray}
\softd_W &=& \text{diag}(w_1,w_2,w_4,w_4,w_4,w_4,w_2,w_1,w_4,w_4,\nn
&& w_4,w_4,w_3,w_3,w_3,w_3,w_3 , w_3, w_3,w_3)\,, \nn
\softd_Z &=& \text{diag}(z_1,z_2,z_7,z_7,z_7,z_7,z_3,z_4,z_8,z_8,\nn
&&z_8,z_8,z_5,z_5,z_6,z_6,z_5,z_5,z_6,z_6)\,, \nn
w_1 &=& T-U+\frac54 i \pi\,, \nn
w_2 &=& -T+U+\frac54 i \pi \,, \nn
w_3 &=&-\frac12(T+U)+\frac34 i \pi\,, \nn
w_4 &=& \frac14 i \pi\,, \nn
z_1 &=& 2 g_{Lu} g_W  (U-T) - i \pi (g_{Lu}^2 +g_W^2)\,, \nn
z_2 &=& 2 g_{Lu} g_W  (T-U) - i \pi (g_{Lu}^2 +g_W^2)\,, \nn
z_3 &=& 2 g_{Ld} g_W  (U-T) - i \pi (g_{Ld}^2 +g_W^2)\,, \nn
z_4 &=& 2 g_{Ld} g_W  (T-U) - i \pi (g_{Ld}^2 +g_W^2)\,, \nn
z_5 &=&  -g_{Ld} g_W T + g_{Lu}g_W U  - i \pi (g_{Lu}g_{Ld} +g_{Lu} g_W -g_{Ld}g_W)\,, \nn
z_6 &=&  g_{Lu} g_W T - g_{Ld }g_W U  - i \pi (g_{Lu}g_{Ld} +g_{Lu} g_W -g_{Ld}g_W)\,, \nn
z_7 &=& -i \pi g_{Lu}^2\,, \nn
z_8 &=& -i \pi g_{Ld}^2\,,
\label{eq208.d}
\end{eqnarray}
and
\begin{eqnarray}
g_W &=& 1-\sin^2\theta_W=\cos^2 \theta_W
\end{eqnarray}
is the $Z$ charge of the $W^+$ boson.

The above equations can also be used to compute radiative corrections to gauge boson pair production by the lepton electroweak doublet, with the obvious replacement of quark $Z$ charges by the corresponding lepton ones, and $\softm_3 \to 0$.

\subsection{Singlets}

Electroweak singlet (right-handed) quarks can produce electroweak gauge bosons. For example gauge boson prodution by right-handed $u$ quarks. The  operator generated at tree-level is
\begin{eqnarray}
O &=& \bar u_{R2} u_{R1} B_4 B_3
\label{285.z}
\end{eqnarray}
with
\begin{eqnarray}
\softm_3 &=& -\sum_{\vev{ij}} \mathbf{T}_i \cdot \mathbf{T}_j\ U_S(n_i,n_j) = \softm^{(3)\,\prime} \,, \nn
\softm_2 &=& -\sum_{\vev{ij}} \mathbf{t}_i \cdot \mathbf{t}_j\ U_S(n_i,n_j) = 0\,, \nn
\softm_1 &=& -\sum_{\vev{ij}} Y_i Y_j\ U_S(n_i,n_j) = \softm^{(1)}\left(0,Y(u_R)\right)\,, \nn
\label{eq186a}
\end{eqnarray}
and the soft anomalous dimension between the scales $Q$ and $\mu_l \sim M_Z$ is given by Eq.~(\ref{softad}), where the matrices $\softm_{(3,2,1)}$ are given by Eqs.~(\ref{eq186a}).

At the scale $\mu_l$, the operator $O$ matches to $\widehat O_i$,
\begin{eqnarray}
\widehat O_1 &=& \bar u_{R2} u_{R1}Z_4 Z_3 \nn
\widehat O_2 &=& \bar u_{R2} u_{R1} A_4 Z_3\nn
\widehat O_3 &=& \bar u_{R2} u_{R1}  Z_4 A_3\nn
\widehat O_4 &=& \bar u_{R2} u_{R1} A_4 A_3
\label{285.a}
\end{eqnarray}
The tree-level matching is 
\begin{eqnarray}
R^{(0)}\left[ \begin{array}{cccc}
\sWsq \\ -\sW \cW \\ - \sW \cW \\ \cWsq
\end{array}
\right]\,,
\end{eqnarray}
and the one-loop soft matching contribution is
\begin{eqnarray}
R_{S,W}^{(1)} &=& 0 \,,\nn
R_{S,Z}^{(1)}  &=& \frac{\alpha_Z}{4\pi}
2 \log\frac{M_Z^2}{\mu^2}\left(-i\pi g_{Ru}^2 \right)R^{(0)}\,.
\label{207}
\end{eqnarray}
The above can also be used for right-handed $d$-quarks with $Y(u_R) \to Y(d_R)$, and for right-handed leptons with $Y(u_R) \to Y(e_R)$ and $\softm_3 \to 0$.

Right-handed $u$ quarks can produce electroweak gauge bosons via
\begin{eqnarray}
O &=& \bar u_{R2} u_{R1} W_4^a W^a_3
\label{286}
\end{eqnarray}
which is not present at tree-level since $W^a$ does not couple to $u_R$. For this operator,
\begin{eqnarray}
\softm_3 &=& -\sum_{\vev{ij}} \mathbf{T}_i \cdot \mathbf{T}_j\ U_S(n_i,n_j) = \softm^{(3)\,\prime} \,, \nn
\softm_2 &=& -\sum_{\vev{ij}} \mathbf{t}_i \cdot \mathbf{t}_j\ U_S(n_i,n_j) = -2 i \pi\,, \nn
\softm_1 &=& -\sum_{\vev{ij}} Y_i Y_j\ U_S(n_i,n_j) = \softm^{(1)}(0,Y(u_R))\,.\nn
\label{eq186c}
\end{eqnarray}

At the low scale, the operator matches to
\begin{eqnarray}
\widehat O_1 &=& \bar u_{R2} u_{R1} W_4^+W_3^- \nn
\widehat O_2 &=& \bar u_{R2} u_{R1} W_4^-W_3^+ \nn
\widehat O_3 &=& \bar u_{R2} u_{R1} Z_4 Z_3 \nn
\widehat O_4 &=& \bar u_{R2} u_{R1}  A_4 Z_3\nn
\widehat O_5 &=& \bar u_{R2} u_{R1} Z_4 A_3 \nn
\widehat O_6 &=& \bar u_{R2} u_{R1} A_4 A_3 
\label{287}
\end{eqnarray}
and the matching condition is $\widehat C = R C$ with tree-level  value
\begin{eqnarray}
R^{(0)} &=& \left[ \begin{array}{c} 1 \\ 1 \\ \cWsq \\ \sW \cW \\ \sW \cW \\ \sWsq \end{array}\right]\,.
\end{eqnarray}

The one-loop matching due to $W$ and $Z$ exchange is given by Eq.~(\ref{167a})
with
\begin{eqnarray}
\softd_W &=&\text{diag} (w_1,w_1,0,0,0,0)\,, \nn
\softd_Z &=& \text{diag}(z_1,z_2,z_3,z_3,z_3,z_3)\,, \nn
w_1 &=& i \pi\,, \nn
z_1 &=& -2 g_{Ru} g_{W}(T-U)-i\pi(g_{Ru}^2+g_W^2)\,, \nn
z_2 &=& 2 g_{Ru} g_{W}(T-U)-i\pi(g_{Ru}^2+g_W^2)\,, \nn
z_3 &=& - i\pi g_{Ru}^2\,.
\end{eqnarray}
The results for $d_R$ and $e_R$ are given by $g_{Ru}\to g_{Rd},g_{Re}$ for the $Z$ charge in $\softd_Z$.

\subsection{Longitudinal bosons via $Q_L \bar Q_L \to \varphi \varphi$}

For longitudinal $W$ production, we also need the results for external unphysical Goldstone boson $\varphi$ fields, which are contained in the Higgs multiplet $\phi$.
The operators are
\begin{eqnarray}
O_1 &=& \bar Q^{(u)} t^a Q^{(u)} \phi^\dagger_4 t^a \phi_3\nn
O_2 &=& \bar Q^{(u)} Q^{(u)} \phi^\dagger_4 \phi_3
\end{eqnarray}
The gauge current $i \left(\phi^\dagger T^a D_\mu \phi - D^\mu \phi^\dagger T^a \phi \right)$ produces operators of this form, weighted by a label  momentum factor $\mathcal{P}_4^\mu-\mathcal{P}_3^\mu$, which is included in the operator coefficients, and is antisymmetric in $3 \leftrightarrow 4$.

The group theory sums needed for the soft anomalous dimension matrix are
\begin{eqnarray}
\softm_3&=& -\sum_{\vev{ij}} \mathbf{T}_i \cdot 
\mathbf{T}_j\ U_S(n_i,n_j)
= \softm^{(3)\,\prime} \otimes \openone \,, \nn
\softm_2&=& -\sum_{\vev{ij}} \mathbf{t}_i \cdot 
\mathbf{t}_j\ U_S(n_i,n_j)= \softm^{(2)}\,, \nn
\softm_1&=&- \sum_{\vev{ij}} Y_i \, Y_j U_S(n_i,n_j)
=  \softm^{(1)}\left(Y(\phi),Y(Q^{(u)})\right) \openone\,. \nn
\end{eqnarray}
At the low scale, the operators match to 20 operators, which have the same structure as the gauge boson operators in Eq.~(\ref{188.a}), with the replacement $W^\pm \to \varphi^\pm$, $W^3 \to \varphi^3$, $B \to H$.
\begin{eqnarray}
\widehat O_1 &=& \bar u_{L2} u_{L1} \varphi^-_4 \varphi^+_3\nn
\widehat O_2 &=& \bar u_{L2} u_{L1} \varphi^3_4  \varphi^3_3 \nn
\widehat O_{3} &=& \bar u_{L2} u_{L1} H_4 \varphi^3_3\nn
\widehat O_{4} &=& \bar u_{L2} u_{L1} \varphi^3_4 H_3\nn
\widehat O_{5} &=& \bar u_{L2} u_{L1} H_4 H_3\nn
\widehat O_6 &=& \bar d_{L2} d_{L1} \varphi^-_4  \varphi^+_3\nn
\widehat O_7 &=& \bar d_{L2} d_{L1} \varphi^3_4 \varphi^3_3 \nn
\widehat O_{8} &=& \bar d_{L2} d_{L1} H_4 \varphi^3_3\nn
\widehat O_{9} &=& \bar d_{L2} d_{L1} \varphi^3_4 H_3\nn
\widehat O_{10} &=& \bar d_{L2} d_{L1} H_4 H_3\nn
\widehat O_{11} &=& \bar u_{L2} d_{L1} \varphi^3_4 \varphi^+_3 \nn
\widehat O_{12} &=& \bar u_{L2} d_{L1} H_4  \varphi^+_3\nn
\widehat O_{13} &=& \bar d_{L2} u_{L1} \varphi^-_4  \varphi^3_3\nn
\widehat O_{14} &=& \bar d_{L2} u_{L1}\varphi^-_4  H_3\,.
\label{188a}
\end{eqnarray}

The convention chosen for the scalar fields is
\begin{eqnarray}
\phi &=& \frac{1}{\sqrt 2}\left[ \begin{array}{c}
\varphi^2 + i \varphi^1 \\
v + H - i \varphi^3 \end{array} \right]\,,
\end{eqnarray}
with $\varphi^\pm = (\varphi^1 \mp i \varphi^2)/\sqrt2$, so that $\varphi^a \propto iT^a \vev{\phi}$. The action of $T_3$ on the neutral fields is
\begin{eqnarray}
T_3 \, H &=& \frac{i}{2} \varphi^3\,,\nn
T_3 \, \varphi^3 &=& -\frac{i}{2} H\,.
\label{t3neutral}
\end{eqnarray}
This causes mixing between $\varphi^3$ and $H$. Under custodial $SU(2)$ symmetry,
the $H$ is a singlet, and $\varphi^3$ belongs to a triplet, so $\varphi^3-H$ mixing is forbidden
by custodial $SU(2)$. In the standard model, custodial $SU(2)$ is violated by hypercharge,
and $\varphi^3-H$ mixing is allowed. In the results derived below, there is $\varphi^3-H$ mixing from $W$ and $Z$ exchange. In the limit $\alpha_W=\alpha_Z$ and $M_W=M_Z$, when custodial $SU(2)$ is restored, the two mixing contributions cancel.

The tree-level matching is
\begin{eqnarray}
R^{(0)} &=& \left[ \begin{array}{cc}  
 \frac 14 & 1 \\
 - \frac18 & \frac12\\
 \frac i 8 & -\frac i 2  \\
 -\frac i 8 & \frac i 2  \\
 - \frac 1 8 & \frac 12 \\
- \frac 14 & 1\\
 \frac 18 & \frac 12  \\
 - \frac i 8 & -\frac i 2  \\
 \frac i 8 & \frac i 2 \\
 \frac 1 8 & \frac 1 2 \\
 - \frac{1}{2 \sqrt 2} & 0\\
\frac{i}{2 \sqrt 2} & 0 \\
-\frac{1}{2 \sqrt 2} & 0\\
-\frac{i}{2 \sqrt 2} & 0\\
\end{array}\right]\,,
\end{eqnarray}
and the one loop matching is given by Eq.~(\ref{167a}) with
\begin{eqnarray}
\softd_W &=& (w_2,\softd_{W1},w_1,\softd_{W2},\softd_{W3},\softd_{W4})\,, \nn
\softd_{W1} &=& \left[\begin{array}{cccc} 
w_0 & w_3 & -w_3 & w_0 \\
-w_3 & w_0 & -w_0 & -w_3 \\
w_3 & -w_0 & w_0  & w_3 \\
w_0 & w_3 & -w_3 & w_0
\end{array}
 \right]\,, \nn
 \softd_{W2} &=& \left[\begin{array}{cccc} 
w_0 & -w_3 & w_3 & w_0 \\
w_3 & w_0 & -w_0 & w_3 \\
-w_3 & -w_0 & w_0  & -w_3 \\
w_0 & -w_3 & w_3 & w_0
\end{array}
 \right]\,, \nn
\softd_{W3} &=& \left[\begin{array}{cc} 
w_4 & i w_4 \\
-i w_4 & w_4 
\end{array}
 \right]\,, \nn
\softd_{W4} &=& \left[\begin{array}{cc} 
w_4 & -i w_4 \\
i w_4 & w_4 
\end{array}
 \right]\,, \nn
w_0 &=& \frac 14 i \pi\,, \nn
w_1 &=& \frac12(T-U)+\frac12 i \pi\,, \nn
w_2 &=& -\frac12(T-U)+\frac12 i \pi\,, \nn
w_3 &=& \frac14i(T-U)\,, \nn
w_4 &=& -\frac14(T+U)+\frac14 i \pi\,, \nn
\softd_Z &=& (z_1,\softd_{Z1},z_2,\softd_{Z2},\softd_{Z3},\softd_{Z4})\,, \nn
\softd_{Z1} &=& \left[\begin{array}{cccc} 
z_3 & -z_4 & z_4 & -z_5 \\
z_4 & z_3 & z_5 & z_4 \\
-z_4 & z_5 & z_3  & -z_4 \\
-z_5 & -z_4 & z_4 & z_3
\end{array}
 \right]\,, \nn
 \softd_{Z2} &=& \left[\begin{array}{cccc} 
z_6 & -z_7 & z_7 & -z_5 \\
z_7 & z_6 & z_5 & z_7 \\
-z_7 & z_5 & z_6  & -z_7 \\
-z_5 & -z_7 & z_7 & z_6
\end{array}
 \right]\,, \nn
 \softd_{Z3} &=&  \left[\begin{array}{cc} 
z_8 & -z_9 \\
z_9 & z_8 
\end{array}
 \right]\,, \nn
\softd_{Z4} &=&  \left[\begin{array}{cc} 
z_8 & z_9 \\
-z_9 & z_8 
\end{array}
 \right]\,, \nn
z_1 &=& 2 g_{\varphi^+} g_{Lu}(T-U)-i\pi(g_{\varphi^+}^2+g_{Lu}^2)\,, \nn
z_2 &=& 2 g_{\varphi^+} g_{Ld}(T-U)-i\pi(g_{\varphi^+}^2+g_{Ld}^2)\,, \nn
z_3 &=& - i \pi g_{Lu}^2\,, \nn
z_4 &=&  \frac12 i g_{Lu}  (T-U)\,, \nn
z_5 &=& \frac14 i \pi\,, \nn
z_6 &=& - i \pi g_{Ld}^2\,, \nn
z_7 &=&  \frac12 i g_{Ld}  (T-U)\,, \nn
z_8 &=& g_{\varphi^+} g_{Lu }T- g_{\varphi^+} g_{Ld}U -i\pi (g_{Lu}g_{Ld}-g_{Ld} g_{\varphi^+}+g_{Lu} g_{\varphi^+})\,, \nn
z_9 &=&  \frac12 i  g_{Ld}T - \frac12 i g_{Lu}U +\frac \pi 2 g_{Ld}-\frac\pi 2 g_{Lu} +\frac\pi 2 g_{\varphi^+}\,.
\end{eqnarray}
The matrices $\softd_{W,Z}$ have block-diagonal form due to $\varphi^3-H$ mixing. The $\varphi^3$ and $\varphi^\pm$ terms are then used to compute longitudinal $Z$ and $W^\pm$ production, using the Goldstone boson equivalence theorem. The equivalence theorem factor $\mathcal{E}$ is included in the collinear function and does not enter the soft matching.

\subsection{Longitudinal bosons via $q_R \bar q_R \to \varphi \varphi$}

Longitudinal gauge bosons are produced by right-handed quarks via operators such as
\begin{eqnarray}
O_1 &=& \bar u_R u_R \phi^\dagger_4 \phi_3
\end{eqnarray}
The group theory sums needed for the soft anomalous dimension matrix are
\begin{eqnarray}
\softm_3&=& -\sum_{\vev{ij}} \mathbf{T}_i \cdot 
\mathbf{T}_j\ U_S(n_i,n_j)
= \softm^{(3)\,\prime}  \,, \nn
\softm_2&=& -\sum_{\vev{ij}} \mathbf{t}_i \cdot 
\mathbf{t}_j\ U_S(n_i,n_j)= \softm^{(2)\,\prime}\,, \nn
\softm_1&=&- \sum_{\vev{ij}} Y_i \, Y_j U_S(n_i,n_j)
=  \softm^{(1)}\left(Y(\phi),Y(u_R)\right) \openone\,.\nn
\end{eqnarray}

At the low scale, the operators match to
\begin{eqnarray}
\widehat O_1 &=& \bar u_{R2} u_{R1}\varphi^{-}_4  \varphi^+_3 \nn
\widehat O_2 &=& \bar u_{R2} u_{R1}\varphi^3_4  \varphi^3_3  \nn
\widehat O_3 &=& \bar u_{R2} u_{R1}H_4  \varphi^3_3 \nn
\widehat O_4 &=& \bar u_{R2} u_{R1} \varphi^3_4 H_3 \nn
\widehat O_5 &=& \bar u_{R2} u_{R1} H_4 H_3 \,.
\end{eqnarray}

The tree-level matching is
 \begin{eqnarray} 
R^{(0)} &=& \left[ \begin{array}{c}
1 \\ \frac12 \\ - \frac i 2 \\ \frac i 2 \\ \frac 12
 \end{array}\right]\,,
 \end{eqnarray}
and the one-loop matching is given by Eq.~(\ref{167a}) with
\begin{eqnarray}
\softd_W &=& \frac14i\pi\, \text{diag}(1,\softd_{W1})\,, \nn
\softd_{W1} &=& \left[ \begin{array}{cccc}
0 & 0 & 0 & 1\\
0 & 0 & -1 & 0\\
0 & -1 & 0 & 0\\
1 & 0 & 0 & 0\\
 \end{array}\right]\,, \nn
\softd_Z &=& (z_1,\softd_{Z1})\,, \nn
\softd_{Z1} &=& \left[ \begin{array}{cccc}
z_3  &  - z_2 & z_2 & -z_4 \\
 z_2 & z_3  & z_4 &   z_2 \\
- z_2 & z_4  &  z_3 & - z_2 \\
-z_4 &   - z_2  &  z_2 & z_3 \\
 \end{array}\right]\,, \nn
z_1 &=& 2 g_{\varphi^+} g_{Ru}(T-U)-i\pi(g_{\varphi^+}^2+g_{Ru}^2)\,, \nn
z_2 &=&  \frac12 ig_{Ru}(T-U)\,, \nn
z_3 &=&  -i \pi g_{Ru}^2\,, \nn
z_4 &=& \frac14 i \pi\,, \nn
g_{\varphi^+} &=& \frac12 - \sin^2 \theta_W\,.
\end{eqnarray}

\section{Soft Functions for Single $W,Z$ Production}\label{sec:singleW}

Single $W$ and $Z$ production proceeds via processes such as $q + \bar q \to W + g$ and $g + q \to W + q$.
The operator basis for production via doublet quarks is
\begin{eqnarray}
O_1 &=& \bar Q^{(u)}_2 T^A t^a Q^{(u)}_1G^A_4  W^a_3
\label{285}
\end{eqnarray}
for the annihilation process,  and
\begin{eqnarray}
O_1 &=& \bar Q^{(u)}_4 T^A t^a Q^{(u)}_1 W^a_3 G^A_2
\label{285b}
\end{eqnarray}
for Compton scattering. The two are related by crossing symmetry.

The matrices for the anomalous dimension are
\begin{eqnarray}
\softm_3 &=& -\sum_{\vev{ij}} \mathbf{T}_i \cdot \mathbf{T}_j\ U_S(n_i,n_j) =
 -\frac{17}{6} i\pi  + \frac32 \left(U+T\right)\,, \nn
\softm_2 &=& -\sum_{\vev{ij}} \mathbf{t}_i \cdot \mathbf{t}_j\ U_S(n_i,n_j) = 
 -\frac74 i\pi  + \left(U+T\right)\,, \nn
\softm_1 &=& -\sum_{\vev{ij}} Y_i Y_j\ U_S(n_i,n_j) = \softm^{(1)}\left(0,Y(Q^{(u)})\right)\,,
\label{eq286.f}
\end{eqnarray}
for annihilation, Eq.~(\ref{285}), and
\begin{eqnarray}
\softm_3 &=& -\sum_{\vev{ij}} \mathbf{T}_i \cdot \mathbf{T}_j\ U_S(n_i,n_j) =
  -\frac{17}{6} i\pi-\frac16 T+\frac32 U\,, \nn
\softm_2 &=& -\sum_{\vev{ij}} \mathbf{t}_i \cdot \mathbf{t}_j\ U_S(n_i,n_j) = 
 -\frac74 i\pi-\frac14 T + U\,, \nn
\softm_1 &=& -\sum_{\vev{ij}} Y_i Y_j\ U_S(n_i,n_j) \nn
& = &\softm^{(1)}\left(Y(Q^{(u)},0,0,Y(Q^{(u)})\right)\,,
\label{eq286.g}
\end{eqnarray}
for Compton scattering, Eq.~(\ref{285b}).

At the low scale $\mu_l \sim M_Z$, the operators Eq.~(\ref{285}) match onto
\begin{eqnarray}
\widehat O_{1} &=& \bar u_{L2}T^A d_{L1} G^A_4  W^+_3\nn
\widehat O_{2} &=& \bar d_{L2}T^A u_{L1}G^A_4  W^-_3 \nn
\widehat O_3 &=& \bar u_{L2} T^A u_{L1} G^A_4 Z_3 \nn
\widehat O_4 &=& \bar u_{L2} T^A u_{L1} G^A_4 A_3 \nn
\widehat O_5 &=& \bar d_{L2} T^A d_{L1} G^A_4  Z_3 \nn
\widehat O_6 &=& \bar d_{L2} T^A d_{L1} G^A_4 A_3 \,.
\label{288.b}
\end{eqnarray}

The tree-level matching for annihilation is
\begin{eqnarray}
 \widehat C_i &=&  \left(R^{(0)}\right)_{ij} C_j \,,\nn
 R^{(0)} &=&
\left[\begin{array}{c} 
\frac{1}{\sqrt2}\\
\frac{1}{\sqrt2}\\
\frac{\cW}{2}\\
\frac{\sW}{2}\\
-\frac{\cW}{2}\\
-\frac{\sW}{2}\\
\end{array}\right]\,,
\label{a111}
\end{eqnarray}
and the one-loop soft matching is given by Eq.~(\ref{167a}) where
\begin{eqnarray}
\softd_W &=& (w_1,w_1,w_2,w_2,w_2,w_2)\,, \nn
\softd_Z &=& (z_1,z_2,z_3,z_3,z_4,z_4)\,, \nn
w_1 &=& -\frac12(T+U)+\frac34 i \pi\,, \nn
w_2 &=& \frac14 i \pi\,, \nn
z_1 &=& g_W g_{Lu} T- g_W g_{Ld} U\nn
&&  -i \pi \left( g_{Ld}g_{Lu} +g_{Lu}g_W-g_{Ld}g_W\right)\,, \nn
z_2 &=& g_W g_{Lu} U- g_W g_{Ld} T \nn
&& -i \pi \left( g_{Ld}g_{Lu} +g_{Lu}g_W-g_{Ld}g_W\right)\,, \nn
z_3 &=&  -i \pi g_{Lu} ^2\,, \nn
z_4 &=&  -i \pi g_{Ld} ^2\,.
\label{eq208.e}
\end{eqnarray}

The results for Compton scattering are given by crossing symmetry. One has to be careful because the collinear functions also need to be transformed. The $\widehat O_i$ operators for Compton scattering are given by swapping the labels $2,4$ in Eq.~(\ref{288}). The tree-level matching remains Eq.~(\ref{a111}), and the one-loop matching is given by Eq.~(\ref{eq208}) with the replacements
\begin{eqnarray}
w_1 &=& \frac14T-\frac12U+\frac34 i \pi\,, \nn
w_2 &=& -\frac14T+\frac14 i \pi\,, \nn
z_1 &=& g_{Ld} g_{Lu} T- g_W g_{Ld} U \nn
&& -i \pi \left( g_{Ld}g_{Lu} +g_{Lu}g_W-g_{Ld}g_W\right)\,, \nn
z_2 &=& g_{Ld} g_{Lu} T + g_W g_{Lu} U \nn
&& -i \pi \left( g_{Ld}g_{Lu} +g_{Lu}g_W-g_{Ld}g_W\right)\,, \nn
z_3 &=&  g_{Lu} ^2\left(T-i \pi\right) \,, \nn
z_4 &=&  g_{Ld} ^2\left(T-i \pi\right)\,.
\label{eq208.f}
\end{eqnarray}

%\subsection{}

The operator basis for single $Z$ production through the $B$ field is
\begin{eqnarray}
O_1 &=& \bar Q^{(u)}_2 T^A  Q^{(u)}_1  G^A_4 B^a_3
\label{285c}
\end{eqnarray}
for annihilation,  and
\begin{eqnarray}
O_1 &=& \bar Q^{(u)}_4 T^A  Q^{(u)}_1 B^a_3 G^A_2
\label{285d}
\end{eqnarray}
for Compton scattering.
In this basis
\begin{eqnarray}
\softm_3 &=& -\sum_{\vev{ij}} \mathbf{T}_i \cdot \mathbf{T}_j\ U_S(n_i,n_j) =
 -\frac43 i\pi  + \frac32 \left(U+T-i\pi\right)\,, \nn
\softm_2 &=& -\sum_{\vev{ij}} \mathbf{t}_i \cdot \mathbf{t}_j\ U_S(n_i,n_j) = 
 -\frac34 i\pi \,, \nn
\softm_1 &=& -\sum_{\vev{ij}} Y_i Y_j\ U_S(n_i,n_j) = \softm^{(1)}\left(0,Y(Q^{(u)})\right)
\label{eq286.a}
\end{eqnarray}
for annihilation, Eq.~(\ref{285c}), and
\begin{eqnarray}
\softm_3 &=& -\sum_{\vev{ij}} \mathbf{T}_i \cdot \mathbf{T}_j\ U_S(n_i,n_j)\nn
& = &
 \frac43(T- i\pi)  + \frac32 \left(U-T-i\pi\right)\,, \nn
\softm_2 &=& -\sum_{\vev{ij}} \mathbf{t}_i \cdot \mathbf{t}_j\ U_S(n_i,n_j) = 
 \frac34 (T-i\pi) \,, \nn
\softm_1 &=& -\sum_{\vev{ij}} Y_i Y_j\ U_S(n_i,n_j)\nn
& = & \softm^{(1)}\left(Y(Q^{(u)},0,0,Y(Q^{(u)})\right)\,, \nn
\label{eq286.b}
\end{eqnarray}
for Compton Scattering, Eq.~(\ref{285d}).

At the low scale $\mu_l \sim M_Z$, the operators Eq.~(\ref{285c}) match onto
\begin{eqnarray}
\widehat O_3 &=& \bar u_{L2} T^A u_{L1} G^A_4 Z_3 \nn
\widehat O_4 &=& \bar u_{L2} T^A u_{L1} G^A_4 A_3 \nn
\widehat O_5 &=& \bar d_{L2} T^A d_{L1} G^A_4 Z_3 \nn
\widehat O_6 &=& \bar d_{L2} T^A d_{L1} G^A_4 A_3 
\label{288}
\end{eqnarray}

The tree-level matching is
\begin{eqnarray}
 \widehat C_i &=&  \left(R^{(0)}\right)_{ij} C_j \,,\nn
 R^{(0)} &=&
\left[\begin{array}{c} 
-\sW \\
\cW \\
-\sW \\
\cW \\
\end{array}\right]\,.
\label{a112}
\end{eqnarray}

The one-loop soft matching is given by Eq.~(\ref{167a}) with
\begin{eqnarray}
\softd_W &=& (w_1,w_1,w_1,w_1) \,, \nn
\softd_Z &=& (z_1,z_1,z_2,z_2)\,, \nn
w_1 &=& \frac14 i \pi\,, \nn
z_1&=&  -i \pi g_{Lu} ^2\,, \nn
z_2 &=&  -i \pi g_{Ld} ^2\,.
\label{eq208a}
\end{eqnarray}

The $\widehat O_i$ operators for Compton scattering are given by swapping the labels $2,4$ in Eq.~(\ref{288}). The tree-level matching remains Eq.~(\ref{a112}), and the one-loop matching is given by Eq.~(\ref{eq208a}) with the replacements
\begin{eqnarray}
\softd_W &=& (w_1,w_1,w_1,w_1) \,, \nn
\softd_Z &=& (z_1,z_1,z_2,z_2)\,, \nn
w_1 &=& -\frac14T+\frac14 i \pi\,, \nn
z_1&=&  g_{Lu} ^2\left(T-i \pi\right)\,, \nn
z_2 &=&  g_{Ld} ^2\left(T-i \pi\right)\,.
\label{eq208a.a}
\end{eqnarray}

%\subsection{}

Single $Z$ production from right-handed quarks proceeds via
\begin{eqnarray}
O &=& \bar u_{R2} T^A u_{R1}G^A_4 B_3 
\label{285f}
\end{eqnarray}
for annihilation, and
\begin{eqnarray}
O &=& \bar u_{R4} T^A u_{R1} B_3 G^A_2
\label{285g}
\end{eqnarray}
for Compton scattering.
The anomalous dimension matrices are
\begin{eqnarray}
\softm_3 &=& -\sum_{\vev{ij}} \mathbf{T}_i \cdot \mathbf{T}_j\ U_S(n_i,n_j) =
 -\frac43 i\pi  + \frac32 \left(U+T-i\pi\right)\,, \nn
\softm_2 &=& -\sum_{\vev{ij}} \mathbf{t}_i \cdot \mathbf{t}_j\ U_S(n_i,n_j) = 
 -\frac34 i\pi\,, \nn
\softm_1 &=& -\sum_{\vev{ij}} Y_i Y_j\ U_S(n_i,n_j) = \softm^{(1)}\left(0,Y(u_R)\right)\,,
\label{eq286.c}
\end{eqnarray}
for annihilation, Eq.~(\ref{285f}), and
\begin{eqnarray}
\softm_3 &=& -\sum_{\vev{ij}} \mathbf{T}_i \cdot \mathbf{T}_j\ U_S(n_i,n_j) =
 \frac43(T- i\pi)  + \frac32 \left(U-T-i\pi\right)\,, \nn
\softm_2 &=& -\sum_{\vev{ij}} \mathbf{t}_i \cdot \mathbf{t}_j\ U_S(n_i,n_j) = 
 \frac34(T- i\pi)\,, \nn
\softm_1 &=& -\sum_{\vev{ij}} Y_i Y_j\ U_S(n_i,n_j) = \softm^{(1)}\left(Y(u_R),0,0,Y(u_R)\right)\,,
\label{eq286.d}
\end{eqnarray}
for Compton scattering, Eq.~(\ref{285g}).

At the low-scale, the operators match to
\begin{eqnarray}
\widehat O_1 &=& \bar u_{R2} T^A u_{R1} G^A_4 Z_3\nn
\widehat O_2 &=& \bar u_{R2} u_{R1} G^A_4 A_3
\label{285.az}
\end{eqnarray}
with tree-level matching
\begin{eqnarray}
R^{(0)}=\left[ \begin{array}{cccc}
-\sW \\ \cW \\
\end{array}\right]\,.
\label{a113}
\end{eqnarray}
The one-loop matching is given by Eq.~(\ref{167a}) with
\begin{eqnarray}
\softd_W &=& 0\,, \nn
\softd_Z &=& (z_1,z_1)\,, \nn
z_1 &=&  -i \pi g_{Ru} ^2\,.
\label{eq208}
\end{eqnarray}

The $\widehat O_i$ operators for Compton scattering are given by swapping the labels $2,4$ in Eq.~(\ref{285}). The tree-level matching remains Eq.~(\ref{a113}), and the one-loop matching is given by Eq.~(\ref{eq208}) with the replacement
\begin{eqnarray}
z_1 &=&  g_{Ru} ^2 (T- i\pi)\,.
\label{eq208.a}
\end{eqnarray}

%FIRST CASE

\section{Soft Functions for Gluon Scattering}\label{sec:gluons}

The operator basis for $q + \bar q \to g + g$ with doublet quarks is
\begin{eqnarray}
O_1 &=& \bar Q^{(u)}_2 Q^{(u)}_1 G^A_4 G^A_3\nn
O_2 &=& \bar Q^{(u)}_2 T^C Q^{(u)}_1d^{ABC} G^A_4 G^B_3\nn
O_3 &=& \bar Q^{(u)}_2 T^C Q^{(u)}_1  i f^{ABC} G^A_4 G^B_3
\label{185.c}
\end{eqnarray}
with soft matrices
\begin{eqnarray}
\softm_3 &=& -\sum_{\vev{ij}} \mathbf{T}_i \cdot \mathbf{T}_j\ U_S(n_i,n_j) =
\softm^{(3,g)}\,, \nn
\softm_2 &=& -\sum_{\vev{ij}} \mathbf{t}_i \cdot \mathbf{t}_j\ U_S(n_i,n_j) = 
\softm^{(2)\prime}\,, \nn
\softm_1 &=& -\sum_{\vev{ij}} Y_i Y_j\ U_S(n_i,n_j) = \softm^{(1)}\left(0,Y(Q^{(u)})\right)\,, \nn
\label{eq286.e}
\end{eqnarray}
This matches onto
\begin{eqnarray}
\widehat O_{11} &=& \bar u_{L2} u_{L1} G^A_4 G^A_3\nn
\widehat O_{12} &=&\bar u_{L2}T^C u_{L1} d^{ABC} G^A_4 G^B_3\nn
\widehat O_{13} &=& \bar u_{L2} T^C u_{L1}  i f^{ABC} G^A_4 G^B_3\nn
\widehat O_{21} &=& \bar d_{L2} d_{L1} G^A_4 G^A_3\nn
\widehat O_{22} &=&\bar d_{L2} T^C d_{L1} d^{ABC} G^A_4 G^B_3\nn
\widehat O_{23} &=& \bar d_{L2} T^C d_{L1}  i f^{ABC} G^A_4 G^B_3
\label{188.b}
\end{eqnarray}
with matching matrix
\begin{eqnarray}
\widehat C_{ia} &=& R_{i} C_{a}\,.
\end{eqnarray}

The tree-level matching is
\begin{eqnarray}
R^{(0)}=\left[ \begin{array}{c}
1 \\ 1
\end{array}\right]\,,
\end{eqnarray}
and the one-loop matching matrices are
\begin{eqnarray}
\softd_W &=& (w_1,w_1)\,, \nn
\softd_Z &=& (z_1,z_2)\,, \nn
w_1 &=& \frac14 i \pi\,, \nn
z_1 &=&  -i \pi g_{Lu} ^2\,, \nn
z_2 &=&  -i \pi g_{Ld} ^2\,.
\label{eq208.b}
\end{eqnarray}

For right-handed quarks, the operator basis is
\begin{eqnarray}
O_1 &=& \bar u_{R2} u_{R1} G^A_4 G^A_3\nn
O_1 &=& \bar u_{R2}T^C  u_{R1} d^{ABC} G^A_4 G^B_3\nn
O_2 &=& \bar u_{R2} T^C u_{R1} i f^{ABC} G^A_4 G^B_3
\label{185.a}
\end{eqnarray}
with
\begin{eqnarray}
\softm_3 &=& -\sum_{\vev{ij}} \mathbf{T}_i \cdot \mathbf{T}_j\ U_S(n_i,n_j) =
\softm^{(3,g)}\,, \nn
\softm_2 &=& -\sum_{\vev{ij}} \mathbf{t}_i \cdot \mathbf{t}_j\ U_S(n_i,n_j) = 0\,, \nn
\softm_1 &=& -\sum_{\vev{ij}} Y_i Y_j\ U_S(n_i,n_j) = \softm^{(1)}\left(0,Y(u_R)\right)\,.\nn
\label{eq286}
\end{eqnarray}

These match onto
\begin{eqnarray}
\widehat O_{1} &=& \bar u_{R2} u_{R1} G^A_4 G^A_3\nn
\widehat O_{2} &=&\bar u_{R2}T^C u_{R1} d^{ABC} G^A_4 G^B_3\nn
\widehat O_{3} &=& \bar u_{R2} T^C u_{R1}  i f^{ABC} G^A_4 G^B_3
\label{188}
\end{eqnarray}

The matching is
\begin{eqnarray}
\widehat C_{a} &=& R C_{a}\,,
\end{eqnarray}
with tree-level matching
\begin{eqnarray}
R^{(0)}=1\,.
\end{eqnarray}
The one-loop matching matrices are
\begin{eqnarray}
\softd_W &=& (0)\,, \nn
\softd_Z &=& (z_1)\,, \nn
z_1 &=&  -i \pi g_{Ru} ^2\,.
\label{eq208.c}
\end{eqnarray}

One can similarly write down the corrections for crossed processes such as $g + q \to g + q$ using crossing, as done above for single electroweak gauge boson production.

\section{Conclusions}\label{sec:conc}

In this paper, we have given the collinear and soft functions needed to compute basic high energy scattering processes in the standard model using the EFT method. The collinear functions have an interesting form, particularly in the weak gauge boson/Higgs sector.

The soft functions can be derived using Eq.~(\ref{152}). They have been explicitly given for a few important cases in this paper. There are many different terms in the scattering operators, because $SU(2) \times U(1)$ and custodial $SU(2)$ are broken in the standard model. The soft anomalous dimensions for QCD have been obtained previously by Kidonakis, Oderda and Sterman~\cite{kidonakis}, and we agree with their results.

Plots of the radiative corrections to various standard model cross-sections of experimental interest, using the results of this paper, have been given in Ref.~\cite{p1}. The radiative corrections give large reductions in the scattering cross-sections at high energy.

\begin{appendix}

\section{Integration of the SCET anomalous dimension}\label{app:integrate}

The analytic formula for integrating the SCET anomalous dimension, with the cusp contribution at three loops, and the non-cusp at two loops, is given here. The result to one lower order was given in Ref.~\cite{Bauer:2003pi}. The collinear anomalous dimension can be integrated using the result below. The soft anomalous dimensions is a matrix, but the matrix structure is $\mu$-independent, so the overall matrix structure is constant at fixed kinematics. Thus it too can be integrated using the results of this appendix, by multiplying the r.h.s. of Eq.~(\ref{soln}) by the constant overall matrix and then taking a matrix exponential.

The anomalous dimension can be written as
\begin{eqnarray}
\gamma(\mu) &=& \left(a A_1 + a^2 A_2 + a^3 A_3 \right) \log \frac{\mu}{\mu_1} + \left(a B_1 + a^2 B_2 \right)\,,\nn
\end{eqnarray}
where $a=\alpha(\mu)/(4\pi)$. The $\beta$-function is
\begin{eqnarray}
\mu \frac{{\rm d}g}{{\rm  d}\mu} &=& - b_0 \frac{g^3}{16\pi^2}- b_1 \frac{g^5}{(16\pi^2)^2}- b_2 \frac{g^7}{(16\pi^2)^3}+\ldots\nn
\end{eqnarray}
Then the solution of
\begin{eqnarray}
\mu \frac{{\rm d} c(\mu) }{{\rm  d}\mu} &=& \gamma(\mu) c(\mu)
\end{eqnarray}
is
\begin{eqnarray}
\frac{c(\mu)}{c(\mu_1)}&=& \exp\left[ \frac{f_0(z)}{\alpha(\mu_1)}+f_1(z)+ \alpha(\mu_1) f_2(z) \right]\,,
\label{soln}
\end{eqnarray}
with
\begin{eqnarray}
z &=& \frac{\alpha(\mu)}{\alpha(\mu_1)}\,, \nn
f_0(z) &=& \frac{\pi A_1}{b_0^2} \left[ \log z +\frac1z-1\right]\,, \nn
f_1(z) &=& \frac{A_1 b_1}{4 b_0^3}\left[\log z -z -\frac12 \log^2 z + 1 \right]-\frac{B_1}{2 b_0} \log z\nn
&& + \frac{A_2}{4 b_0^2}\left[z-\log z - 1\right]\,, \nn
f_2(z) &=&\frac{A_1 b_1^2}{32 \pi b_0^4} \left[z^2-2z+2z \log z -2 \log z + 1\right]\nn
&&+\frac{A_1 b_2}{32 \pi b_0^3}\left[2\log z-z^2+1 \right]\nn
&&-\frac{A_2 b_1}{32 \pi b_0^3} \left[z^2+2 z \log z -4 z + 3\right] \nn
&&+ \frac{A_3}{32 \pi b_0^2}\left[z^2-2z+1\right] +
\left[\frac{B_1 b_1}{8 \pi b_0^2}-\frac{B_2}{8 \pi b_0} \right] \left[z-1\right]\,.\nn
\end{eqnarray}

\section{Wavefunction Factors}\label{app:wave}

The transverse gauge boson inverse-propagator is
\begin{eqnarray}
-i\left(g_{\mu \nu}-\frac{k_\mu k_\nu}{k^2}\right)\left[ \begin{array}{cc} 
k^2-M_Z^2-\Pi_{ZZ}(k^2) & -\Pi_{Z\gamma}(k^2)\\
-\Pi_{\gamma Z}(k^2) & k^2-\Pi_{\gamma\gamma}(k^2)
\end{array}\right]\,,\nn
\end{eqnarray}
where $\Pi=0$ at tree-level, and $M_Z$ is the tree-level $Z$-boson mass.
Then the wavefunction factors to one-loop are
\begin{eqnarray}
\delta \waver_Z &=& \Pi_{ZZ}^\prime(M_Z^2)\,,\nn
\delta \waver_\gamma &=& \Pi_{\gamma\gamma}^\prime(0)\,,\nn
\waver_{\gamma \to Z} &=& \frac{1}{M_Z^2} \Pi_{Z\gamma}^\prime(M_Z^2)\,,\nn
\waver_{Z \to \gamma} &=& -\frac{1}{M_Z^2} \Pi_{\gamma Z}^\prime(0)\,.
\end{eqnarray}

\begin{widetext}
\section{Radiative Corrections to the Equivalence Theorem}\label{app:et}

The equivalence theorem radiative correction factor $\mathcal{E}$ (defined as in Ref.~\cite{p1}) for longitudinal $W$ and $Z$ production is $\mathcal{E}_{W,Z} = 1+ \mathcal{E}^{(1)}_{W,Z} \aem/(4 \pi \sin^2 \theta_W)$. The one-loop corrections in $R_{\xi=1}$ gauge are
\begin{eqnarray}
\mathcal{E}_{W}^{(1)} &=& \frac{m_t^2}{2 M_W^2}-\frac{M_H^2}{12 M_W^2}-\frac{M_Z^2}{12 M_W^2} - \frac 32 
-\left(\frac{M_H^2}{12 M_W^2}+\frac{M_Z^2}{12 M_W^2}+\frac23 \right) \frac{A_0(M_W^2)}{M_W^2}
+\left(\frac{M_Z^4}{12 M_W^4}+\frac{M_Z^2}{2 M_W^2}-\frac43 \right) \frac{A_0(M_Z^2)}{M_Z^2}\nn
&&+\left(\frac{M_H^4}{12 M_W^4}-\frac{M_H^2}{6 M_W^2} \right) \frac{A_0(M_H^2)}{M_H^2}
+\left(\frac{m_t^2}{2 M_W^2}-\frac{m_t^4}{2 M_W^4} \right) \frac{A_0(m_t^2)}{m_t^2}
+\left(\frac{M_Z^4}{12 M_W^4}+\frac{5M_Z^2}{12 M_W^2}-2 \right) B_0(-M_W^2,M_Z,M_W)\nn
&&+\left(\frac{M_H^4}{12 M_W^4}-\frac{M_H^2}{4 M_W^2} \right) B_0(-M_W^2,M_W,M_H)
+\left(\frac{m_t^2}{2 M_W^2}-\frac{m_t^4}{2 M_W^4} \right)B_0(-M_W^2,0,m_t)
+\frac{3 M_W^2}{2} B_0^\prime (-M_W^2,0,0)\nn
&&+\left( \frac{2M_W^4}{M_Z^2}-2 M_W^2\right) B_0^\prime(-M_W^2,0,M_W)
-\left( \frac{2M_W^4}{M_Z^2}+\frac{41M_W^2 }{24}-\frac{M_Z^2}{6}+
\frac{M_Z^4}{12 M_W^2}\right) B_0^\prime(-M_W^2,M_Z,M_W)\nn
&&+\left( -\frac{M_H^4}{12 M_W^2}+\frac{M_H^2}{12}+\frac{5 M_W^2}{8}\right) B_0^\prime(-M_W^2,M_W,M_H)
+\left(\frac{m_t^4}{2 M_W^2}-m_t^2+\frac{M_W^2}{2} \right)B_0^\prime(-M_W^2,0,m_t)\,,\nn
\mathcal{E}_{Z}^{(1)} &=& \frac{17 m_t^2}{18 M_W^2}-\frac{20 m_t^2}{9 M_Z^2}+\frac{16 M_W^2 m_t^2}{9 M_Z^4}
-\frac{M_H^2}{12 M_W^2}-\frac{M_Z^2}{12 M_W^2} - \frac 16+\frac{2 M_W^2}{3M_Z^2}-\frac{2 M_W^4}{M_Z^4}
-\left(\frac{2 M_W^4}{M_Z^4}-\frac{2 M_W^2}{3 M_Z^2}+\frac16 \right) \frac{A_0(M_W^2)}{M_W^2}\nn
&&-\frac{M_H^2}{12M_W^2} \frac{A_0(M_Z^2)}{M_Z^2}+\left(\frac{M_H^4}{12 M_W^2 M_Z^2}-\frac{M_H^2}{6 M_W^2} \right) \frac{A_0(M_H^2)}{M_H^2}
+\left(\frac{17m_t^2}{18 M_W^2}-\frac{20 m_t^2}{9 M_Z^2}+\frac{16 M_W^2 m_t^2}{9 M_Z^4} \right) \frac{A_0(m_t^2)}{m_t^2}\nn
&&+\left(-\frac{2M_W^4}{M_Z^4}+\frac{2 M_W^2}{3 M_Z^2}-\frac16 \right) B_0(-M_Z^2,M_W,M_W)+\left(\frac{M_H^4}{12 M_W^2 M_Z^2}-\frac{M_H^2}{4 M_W^2} \right) B_0(-M_Z^2,M_Z,M_H)\nn
&&+\left(\frac{17 m_t^2}{18 M_W^2}-\frac{20 m_t^2}{9 M_Z^2} +\frac{16 M_W^2 m_t^2}{9 M_Z^4}\right)
B_0(-M_Z^2,m_t,m_t)
+\left(\frac{103 M_Z^4}{36 M_W^2}-\frac{50 M_Z^2}{9}+\frac{40 M_W^2}{9} \right) B_0^\prime (-M_Z^2,0,0)\nn
&&+\left(-\frac{M_H^4}{12M_W^2}+\frac{M_Z^2M_H^2}{12M_W^2}+\frac{5 M_Z^4}{8M_W^2} \right)B_0^\prime(-M_Z^2,M_Z,M_H)
%%%%
+\left(-\frac{2M_W^4}{M_Z^2}-\frac{17 M_W^2}{6}+\frac{7 M_Z^2}{6}+\frac{M_Z^4}{24 M_W^2}\right)B_0^\prime(-M_Z^2,M_W,M_W)\nn
&&+\left(\frac{17 M_Z^4}{36 M_W^2}-\frac{5 m_t^2 M_Z^2}{9M_W^2}
-\frac{10 M_Z^2}{9}+\frac{8 M_W^2}{9}-\frac{20 m_t^2}{9}+\frac{16 M_W^2 m_t^2}{9 M_Z^2} \right) B_0^\prime(-M_Z^2,m_t,m_t)\,,
\label{eqC1}
\end{eqnarray}
where $A_0,B_0,B_0^\prime$ are given in Eqs.~(\ref{eq2})--(\ref{B0def}) using the conventions of Ref.~\cite{bardin}. The $A_0$ and $B_0$ functions are ultraviolet divergent,
\begin{eqnarray}
\frac{A_0(m^2)}{m^2} = - \frac1{\eUV} + \text{UV finite}, \qquad B_0(p^2,m_1,m_2) =  \frac1{\eUV} + \text{UV finite}\,,
\end{eqnarray}
and the infrared divergent function is
\begin{eqnarray}
B_0^\prime(-M_W^2,0,M_W) &=& \frac{1}{\eIR} \frac{1}{2M_W^2} + \text{IR finite}\,.
\end{eqnarray}
$\mathcal{E}_{W,Z}^{(1)}$ are ultraviolet finite, and $\mathcal{E}_{Z}^{(1)}$ is infrared finite. The infrared divergence in 
$\mathcal{E}_{W}^{(1)}$ is
\begin{eqnarray}
\mathcal{E}_{W}^{(1)} &=&\frac{1}{\eIR} \left( \frac{M_W^2}{M_Z^2}-1\right) + \text{IR finite}
\end{eqnarray}
and is proportional to $1-M_W^2/M_Z^2 = \sin^2 \theta_W$, which indicates that it arises from photon exchange. $\mathcal{E}_W$ in Eq.~(\ref{eq34}) is treated as a matching condition (see footnote~\ref{foot}), i.e.\ the $1/\eIR$ terms in Eq.~(\ref{eqC1}) are dropped. This procedure is valid provided the infrared divergences of the original theory agree with those of the effective theory, so that the $1/\eIR$ terms cancel  in the matching condition.

\end{widetext}

\end{appendix}

\bibliography{biblio}

\begin{thebibliography}{65}
\expandafter\ifx\csname natexlab\endcsname\relax\def\natexlab#1{#1}\fi
\expandafter\ifx\csname bibnamefont\endcsname\relax
  \def\bibnamefont#1{#1}\fi
\expandafter\ifx\csname bibfnamefont\endcsname\relax
  \def\bibfnamefont#1{#1}\fi
\expandafter\ifx\csname citenamefont\endcsname\relax
  \def\citenamefont#1{#1}\fi
\expandafter\ifx\csname url\endcsname\relax
  \def\url#1{\texttt{#1}}\fi
\expandafter\ifx\csname urlprefix\endcsname\relax\def\urlprefix{URL }\fi
\providecommand{\bibinfo}[2]{#2}
\providecommand{\eprint}[2][]{\url{#2}}

\bibitem[{\citenamefont{Bauer et~al.}(2000)\citenamefont{Bauer, Fleming, and
  Luke}}]{BFL}
\bibinfo{author}{\bibfnamefont{C.~W.} \bibnamefont{Bauer}},
  \bibinfo{author}{\bibfnamefont{S.}~\bibnamefont{Fleming}}, \bibnamefont{and}
  \bibinfo{author}{\bibfnamefont{M.~E.} \bibnamefont{Luke}},
  \bibinfo{journal}{Phys. Rev.} \textbf{\bibinfo{volume}{D63}},
  \bibinfo{pages}{014006} (\bibinfo{year}{2000}), \eprint{hep-ph/0005275}.

\bibitem[{\citenamefont{Bauer et~al.}(2001)\citenamefont{Bauer, Fleming,
  Pirjol, and Stewart}}]{SCET1}
\bibinfo{author}{\bibfnamefont{C.~W.} \bibnamefont{Bauer}},
  \bibinfo{author}{\bibfnamefont{S.}~\bibnamefont{Fleming}},
  \bibinfo{author}{\bibfnamefont{D.}~\bibnamefont{Pirjol}}, \bibnamefont{and}
  \bibinfo{author}{\bibfnamefont{I.~W.} \bibnamefont{Stewart}},
  \bibinfo{journal}{Phys. Rev.} \textbf{\bibinfo{volume}{D63}},
  \bibinfo{pages}{114020} (\bibinfo{year}{2001}), \eprint{hep-ph/0011336}.

\bibitem[{\citenamefont{Bauer and Stewart}(2001)}]{SCET2}
\bibinfo{author}{\bibfnamefont{C.~W.} \bibnamefont{Bauer}} \bibnamefont{and}
  \bibinfo{author}{\bibfnamefont{I.~W.} \bibnamefont{Stewart}},
  \bibinfo{journal}{Phys. Lett.} \textbf{\bibinfo{volume}{B516}},
  \bibinfo{pages}{134} (\bibinfo{year}{2001}), \eprint{hep-ph/0107001}.

\bibitem[{\citenamefont{Bauer et~al.}(2002{\natexlab{a}})\citenamefont{Bauer,
  Pirjol, and Stewart}}]{BPS}
\bibinfo{author}{\bibfnamefont{C.~W.} \bibnamefont{Bauer}},
  \bibinfo{author}{\bibfnamefont{D.}~\bibnamefont{Pirjol}}, \bibnamefont{and}
  \bibinfo{author}{\bibfnamefont{I.~W.} \bibnamefont{Stewart}},
  \bibinfo{journal}{Phys. Rev.} \textbf{\bibinfo{volume}{D65}},
  \bibinfo{pages}{054022} (\bibinfo{year}{2002}{\natexlab{a}}),
  \eprint{hep-ph/0109045}.

\bibitem[{\citenamefont{Chiu et~al.}(2008{\natexlab{a}})\citenamefont{Chiu,
  Golf, Kelley, and Manohar}}]{CGKM1}
\bibinfo{author}{\bibfnamefont{J.-Y.} \bibnamefont{Chiu}},
  \bibinfo{author}{\bibfnamefont{F.}~\bibnamefont{Golf}},
  \bibinfo{author}{\bibfnamefont{R.}~\bibnamefont{Kelley}}, \bibnamefont{and}
  \bibinfo{author}{\bibfnamefont{A.~V.} \bibnamefont{Manohar}},
  \bibinfo{journal}{Phys. Rev. Lett.} \textbf{\bibinfo{volume}{100}},
  \bibinfo{pages}{021802} (\bibinfo{year}{2008}{\natexlab{a}}),
  \eprint{0709.2377}.

\bibitem[{\citenamefont{Chiu et~al.}(2008{\natexlab{b}})\citenamefont{Chiu,
  Golf, Kelley, and Manohar}}]{CGKM2}
\bibinfo{author}{\bibfnamefont{J.-Y.} \bibnamefont{Chiu}},
  \bibinfo{author}{\bibfnamefont{F.}~\bibnamefont{Golf}},
  \bibinfo{author}{\bibfnamefont{R.}~\bibnamefont{Kelley}}, \bibnamefont{and}
  \bibinfo{author}{\bibfnamefont{A.~V.} \bibnamefont{Manohar}},
  \bibinfo{journal}{Phys. Rev.} \textbf{\bibinfo{volume}{D77}},
  \bibinfo{pages}{053004} (\bibinfo{year}{2008}{\natexlab{b}}),
  \eprint{0712.0396}.

\bibitem[{\citenamefont{Chiu et~al.}(2008{\natexlab{c}})\citenamefont{Chiu,
  Kelley, and Manohar}}]{CKM}
\bibinfo{author}{\bibfnamefont{J.-Y.} \bibnamefont{Chiu}},
  \bibinfo{author}{\bibfnamefont{R.}~\bibnamefont{Kelley}}, \bibnamefont{and}
  \bibinfo{author}{\bibfnamefont{A.~V.} \bibnamefont{Manohar}},
  \bibinfo{journal}{Phys. Rev.} \textbf{\bibinfo{volume}{D78}},
  \bibinfo{pages}{073006} (\bibinfo{year}{2008}{\natexlab{c}}),
  \eprint{0806.1240}.

\bibitem[{\citenamefont{Chiu et~al.}(2009{\natexlab{a}})\citenamefont{Chiu,
  Fuhrer, Kelley, and Manohar}}]{p1}
\bibinfo{author}{\bibfnamefont{J.~Y.} \bibnamefont{Chiu}},
  \bibinfo{author}{\bibfnamefont{A.}~\bibnamefont{Fuhrer}},
  \bibinfo{author}{\bibfnamefont{R.}~\bibnamefont{Kelley}}, \bibnamefont{and}
  \bibinfo{author}{\bibfnamefont{A.~V.} \bibnamefont{Manohar}}
  (\bibinfo{year}{2009}{\natexlab{a}}), \eprint{0909.0012}.

\bibitem[{\citenamefont{Ciafaloni et~al.}(2000)\citenamefont{Ciafaloni,
  Ciafaloni, and Comelli}}]{ccc}
\bibinfo{author}{\bibfnamefont{M.}~\bibnamefont{Ciafaloni}},
  \bibinfo{author}{\bibfnamefont{P.}~\bibnamefont{Ciafaloni}},
  \bibnamefont{and} \bibinfo{author}{\bibfnamefont{D.}~\bibnamefont{Comelli}},
  \bibinfo{journal}{Phys. Rev. Lett.} \textbf{\bibinfo{volume}{84}},
  \bibinfo{pages}{4810} (\bibinfo{year}{2000}), \eprint{hep-ph/0001142}.

\bibitem[{\citenamefont{Ciafaloni and Comelli}(1999)}]{ciafaloni1}
\bibinfo{author}{\bibfnamefont{P.}~\bibnamefont{Ciafaloni}} \bibnamefont{and}
  \bibinfo{author}{\bibfnamefont{D.}~\bibnamefont{Comelli}},
  \bibinfo{journal}{Phys. Lett.} \textbf{\bibinfo{volume}{B446}},
  \bibinfo{pages}{278} (\bibinfo{year}{1999}), \eprint{hep-ph/9809321}.

\bibitem[{\citenamefont{Ciafaloni and Comelli}(2000)}]{ciafaloni2}
\bibinfo{author}{\bibfnamefont{P.}~\bibnamefont{Ciafaloni}} \bibnamefont{and}
  \bibinfo{author}{\bibfnamefont{D.}~\bibnamefont{Comelli}},
  \bibinfo{journal}{Phys. Lett.} \textbf{\bibinfo{volume}{B476}},
  \bibinfo{pages}{49} (\bibinfo{year}{2000}), \eprint{hep-ph/9910278}.

\bibitem[{\citenamefont{Fadin et~al.}(2000)\citenamefont{Fadin, Lipatov,
  Martin, and Melles}}]{fadin}
\bibinfo{author}{\bibfnamefont{V.~S.} \bibnamefont{Fadin}},
  \bibinfo{author}{\bibfnamefont{L.~N.} \bibnamefont{Lipatov}},
  \bibinfo{author}{\bibfnamefont{A.~D.} \bibnamefont{Martin}},
  \bibnamefont{and} \bibinfo{author}{\bibfnamefont{M.}~\bibnamefont{Melles}},
  \bibinfo{journal}{Phys. Rev.} \textbf{\bibinfo{volume}{D61}},
  \bibinfo{pages}{094002} (\bibinfo{year}{2000}), \eprint{hep-ph/9910338}.

\bibitem[{\citenamefont{Kuhn et~al.}(2000)\citenamefont{Kuhn, Penin, and
  Smirnov}}]{kps}
\bibinfo{author}{\bibfnamefont{J.~H.} \bibnamefont{Kuhn}},
  \bibinfo{author}{\bibfnamefont{A.~A.} \bibnamefont{Penin}}, \bibnamefont{and}
  \bibinfo{author}{\bibfnamefont{V.~A.} \bibnamefont{Smirnov}},
  \bibinfo{journal}{Eur. Phys. J.} \textbf{\bibinfo{volume}{C17}},
  \bibinfo{pages}{97} (\bibinfo{year}{2000}), \eprint{hep-ph/9912503}.

\bibitem[{\citenamefont{Feucht et~al.}(2004)\citenamefont{Feucht, Kuhn, Penin,
  and Smirnov}}]{fkps}
\bibinfo{author}{\bibfnamefont{B.}~\bibnamefont{Feucht}},
  \bibinfo{author}{\bibfnamefont{J.~H.} \bibnamefont{Kuhn}},
  \bibinfo{author}{\bibfnamefont{A.~A.} \bibnamefont{Penin}}, \bibnamefont{and}
  \bibinfo{author}{\bibfnamefont{V.~A.} \bibnamefont{Smirnov}},
  \bibinfo{journal}{Phys. Rev. Lett.} \textbf{\bibinfo{volume}{93}},
  \bibinfo{pages}{101802} (\bibinfo{year}{2004}), \eprint{hep-ph/0404082}.

\bibitem[{\citenamefont{Jantzen
  et~al.}(2005{\natexlab{a}})\citenamefont{Jantzen, Kuhn, Penin, and
  Smirnov}}]{jkps}
\bibinfo{author}{\bibfnamefont{B.}~\bibnamefont{Jantzen}},
  \bibinfo{author}{\bibfnamefont{J.~H.} \bibnamefont{Kuhn}},
  \bibinfo{author}{\bibfnamefont{A.~A.} \bibnamefont{Penin}}, \bibnamefont{and}
  \bibinfo{author}{\bibfnamefont{V.~A.} \bibnamefont{Smirnov}},
  \bibinfo{journal}{Phys. Rev.} \textbf{\bibinfo{volume}{D72}},
  \bibinfo{pages}{051301} (\bibinfo{year}{2005}{\natexlab{a}}),
  \eprint{hep-ph/0504111}.

\bibitem[{\citenamefont{Jantzen
  et~al.}(2005{\natexlab{b}})\citenamefont{Jantzen, Kuhn, Penin, and
  Smirnov}}]{jkps4}
\bibinfo{author}{\bibfnamefont{B.}~\bibnamefont{Jantzen}},
  \bibinfo{author}{\bibfnamefont{J.~H.} \bibnamefont{Kuhn}},
  \bibinfo{author}{\bibfnamefont{A.~A.} \bibnamefont{Penin}}, \bibnamefont{and}
  \bibinfo{author}{\bibfnamefont{V.~A.} \bibnamefont{Smirnov}},
  \bibinfo{journal}{Nucl. Phys.} \textbf{\bibinfo{volume}{B731}},
  \bibinfo{pages}{188} (\bibinfo{year}{2005}{\natexlab{b}}),
  \eprint{hep-ph/0509157}.

\bibitem[{\citenamefont{Beccaria et~al.}(2001)\citenamefont{Beccaria, Renard,
  and Verzegnassi}}]{beccaria}
\bibinfo{author}{\bibfnamefont{M.}~\bibnamefont{Beccaria}},
  \bibinfo{author}{\bibfnamefont{F.~M.} \bibnamefont{Renard}},
  \bibnamefont{and}
  \bibinfo{author}{\bibfnamefont{C.}~\bibnamefont{Verzegnassi}},
  \bibinfo{journal}{Phys. Rev.} \textbf{\bibinfo{volume}{D63}},
  \bibinfo{pages}{053013} (\bibinfo{year}{2001}), \eprint{hep-ph/0010205}.

\bibitem[{\citenamefont{Denner and Pozzorini}(2001{\natexlab{a}})}]{dp1}
\bibinfo{author}{\bibfnamefont{A.}~\bibnamefont{Denner}} \bibnamefont{and}
  \bibinfo{author}{\bibfnamefont{S.}~\bibnamefont{Pozzorini}},
  \bibinfo{journal}{Eur. Phys. J.} \textbf{\bibinfo{volume}{C18}},
  \bibinfo{pages}{461} (\bibinfo{year}{2001}{\natexlab{a}}),
  \eprint{hep-ph/0010201}.

\bibitem[{\citenamefont{Denner and Pozzorini}(2001{\natexlab{b}})}]{dp2}
\bibinfo{author}{\bibfnamefont{A.}~\bibnamefont{Denner}} \bibnamefont{and}
  \bibinfo{author}{\bibfnamefont{S.}~\bibnamefont{Pozzorini}},
  \bibinfo{journal}{Eur. Phys. J.} \textbf{\bibinfo{volume}{C21}},
  \bibinfo{pages}{63} (\bibinfo{year}{2001}{\natexlab{b}}),
  \eprint{hep-ph/0104127}.

\bibitem[{\citenamefont{Hori et~al.}(2000)\citenamefont{Hori, Kawamura, and
  Kodaira}}]{hori}
\bibinfo{author}{\bibfnamefont{M.}~\bibnamefont{Hori}},
  \bibinfo{author}{\bibfnamefont{H.}~\bibnamefont{Kawamura}}, \bibnamefont{and}
  \bibinfo{author}{\bibfnamefont{J.}~\bibnamefont{Kodaira}},
  \bibinfo{journal}{Phys. Lett.} \textbf{\bibinfo{volume}{B491}},
  \bibinfo{pages}{275} (\bibinfo{year}{2000}), \eprint{hep-ph/0007329}.

\bibitem[{\citenamefont{Beenakker and Werthenbach}(2002)}]{beenakker}
\bibinfo{author}{\bibfnamefont{W.}~\bibnamefont{Beenakker}} \bibnamefont{and}
  \bibinfo{author}{\bibfnamefont{A.}~\bibnamefont{Werthenbach}},
  \bibinfo{journal}{Nucl. Phys.} \textbf{\bibinfo{volume}{B630}},
  \bibinfo{pages}{3} (\bibinfo{year}{2002}), \eprint{hep-ph/0112030}.

\bibitem[{\citenamefont{Denner et~al.}(2003)\citenamefont{Denner, Melles, and
  Pozzorini}}]{dmp}
\bibinfo{author}{\bibfnamefont{A.}~\bibnamefont{Denner}},
  \bibinfo{author}{\bibfnamefont{M.}~\bibnamefont{Melles}}, \bibnamefont{and}
  \bibinfo{author}{\bibfnamefont{S.}~\bibnamefont{Pozzorini}},
  \bibinfo{journal}{Nucl. Phys.} \textbf{\bibinfo{volume}{B662}},
  \bibinfo{pages}{299} (\bibinfo{year}{2003}), \eprint{hep-ph/0301241}.

\bibitem[{\citenamefont{Pozzorini}(2004)}]{pozzorini}
\bibinfo{author}{\bibfnamefont{S.}~\bibnamefont{Pozzorini}},
  \bibinfo{journal}{Nucl. Phys.} \textbf{\bibinfo{volume}{B692}},
  \bibinfo{pages}{135} (\bibinfo{year}{2004}), \eprint{hep-ph/0401087}.

\bibitem[{\citenamefont{Jantzen and Smirnov}(2006)}]{js}
\bibinfo{author}{\bibfnamefont{B.}~\bibnamefont{Jantzen}} \bibnamefont{and}
  \bibinfo{author}{\bibfnamefont{V.~A.} \bibnamefont{Smirnov}},
  \bibinfo{journal}{Eur. Phys. J.} \textbf{\bibinfo{volume}{C47}},
  \bibinfo{pages}{671} (\bibinfo{year}{2006}), \eprint{hep-ph/0603133}.

\bibitem[{\citenamefont{Melles}(2000)}]{melles1}
\bibinfo{author}{\bibfnamefont{M.}~\bibnamefont{Melles}},
  \bibinfo{journal}{Phys. Lett.} \textbf{\bibinfo{volume}{B495}},
  \bibinfo{pages}{81} (\bibinfo{year}{2000}), \eprint{hep-ph/0006077}.

\bibitem[{\citenamefont{Melles}(2001{\natexlab{a}})}]{melles2}
\bibinfo{author}{\bibfnamefont{M.}~\bibnamefont{Melles}},
  \bibinfo{journal}{Phys. Rev.} \textbf{\bibinfo{volume}{D63}},
  \bibinfo{pages}{034003} (\bibinfo{year}{2001}{\natexlab{a}}),
  \eprint{hep-ph/0004056}.

\bibitem[{\citenamefont{Melles}(2003)}]{melles3}
\bibinfo{author}{\bibfnamefont{M.}~\bibnamefont{Melles}},
  \bibinfo{journal}{Phys. Rept.} \textbf{\bibinfo{volume}{375}},
  \bibinfo{pages}{219} (\bibinfo{year}{2003}), \eprint{hep-ph/0104232}.

\bibitem[{\citenamefont{Kuhn et~al.}(2008)\citenamefont{Kuhn, Metzler, and
  Penin}}]{kuhnW}
\bibinfo{author}{\bibfnamefont{J.~H.} \bibnamefont{Kuhn}},
  \bibinfo{author}{\bibfnamefont{F.}~\bibnamefont{Metzler}}, \bibnamefont{and}
  \bibinfo{author}{\bibfnamefont{A.~A.} \bibnamefont{Penin}},
  \bibinfo{journal}{Nucl. Phys.} \textbf{\bibinfo{volume}{B795}},
  \bibinfo{pages}{277} (\bibinfo{year}{2008}), \eprint{0709.4055}.

\bibitem[{\citenamefont{Denner et~al.}(2007)\citenamefont{Denner, Jantzen, and
  Pozzorini}}]{Denner:2006jr}
\bibinfo{author}{\bibfnamefont{A.}~\bibnamefont{Denner}},
  \bibinfo{author}{\bibfnamefont{B.}~\bibnamefont{Jantzen}}, \bibnamefont{and}
  \bibinfo{author}{\bibfnamefont{S.}~\bibnamefont{Pozzorini}},
  \bibinfo{journal}{Nucl. Phys.} \textbf{\bibinfo{volume}{B761}},
  \bibinfo{pages}{1} (\bibinfo{year}{2007}), \eprint{hep-ph/0608326}.

\bibitem[{\citenamefont{Denner et~al.}(2008)\citenamefont{Denner, Jantzen, and
  Pozzorini}}]{Denner:2008yn}
\bibinfo{author}{\bibfnamefont{A.}~\bibnamefont{Denner}},
  \bibinfo{author}{\bibfnamefont{B.}~\bibnamefont{Jantzen}}, \bibnamefont{and}
  \bibinfo{author}{\bibfnamefont{S.}~\bibnamefont{Pozzorini}},
  \bibinfo{journal}{JHEP} \textbf{\bibinfo{volume}{11}}, \bibinfo{pages}{062}
  (\bibinfo{year}{2008}), \eprint{0809.0800}.

\bibitem[{\citenamefont{Melles}(2001{\natexlab{b}})}]{melles4}
\bibinfo{author}{\bibfnamefont{M.}~\bibnamefont{Melles}},
  \bibinfo{journal}{Phys. Rev.} \textbf{\bibinfo{volume}{D64}},
  \bibinfo{pages}{014011} (\bibinfo{year}{2001}{\natexlab{b}}),
  \eprint{hep-ph/0012157}.

\bibitem[{\citenamefont{Kidonakis et~al.}(1998)\citenamefont{Kidonakis, Oderda,
  and Sterman}}]{kidonakis}
\bibinfo{author}{\bibfnamefont{N.}~\bibnamefont{Kidonakis}},
  \bibinfo{author}{\bibfnamefont{G.}~\bibnamefont{Oderda}}, \bibnamefont{and}
  \bibinfo{author}{\bibfnamefont{G.}~\bibnamefont{Sterman}},
  \bibinfo{journal}{Nucl. Phys.} \textbf{\bibinfo{volume}{B531}},
  \bibinfo{pages}{365} (\bibinfo{year}{1998}), \eprint{hep-ph/9803241}.

\bibitem[{\citenamefont{Luke and Manohar}(1992)}]{rpi1}
\bibinfo{author}{\bibfnamefont{M.~E.} \bibnamefont{Luke}} \bibnamefont{and}
  \bibinfo{author}{\bibfnamefont{A.~V.} \bibnamefont{Manohar}},
  \bibinfo{journal}{Phys. Lett.} \textbf{\bibinfo{volume}{B286}},
  \bibinfo{pages}{348} (\bibinfo{year}{1992}), \eprint{hep-ph/9205228}.

\bibitem[{\citenamefont{Manohar et~al.}(2002)\citenamefont{Manohar, Mehen,
  Pirjol, and Stewart}}]{rpi2}
\bibinfo{author}{\bibfnamefont{A.~V.} \bibnamefont{Manohar}},
  \bibinfo{author}{\bibfnamefont{T.}~\bibnamefont{Mehen}},
  \bibinfo{author}{\bibfnamefont{D.}~\bibnamefont{Pirjol}}, \bibnamefont{and}
  \bibinfo{author}{\bibfnamefont{I.~W.} \bibnamefont{Stewart}},
  \bibinfo{journal}{Phys. Lett.} \textbf{\bibinfo{volume}{B539}},
  \bibinfo{pages}{59} (\bibinfo{year}{2002}), \eprint{hep-ph/0204229}.

\bibitem[{\citenamefont{Manohar}(2003)}]{dis}
\bibinfo{author}{\bibfnamefont{A.~V.} \bibnamefont{Manohar}},
  \bibinfo{journal}{Phys. Rev.} \textbf{\bibinfo{volume}{D68}},
  \bibinfo{pages}{114019} (\bibinfo{year}{2003}).

\bibitem[{\citenamefont{Collins and Hautmann}(2000)}]{hautmann}
\bibinfo{author}{\bibfnamefont{J.~C.} \bibnamefont{Collins}} \bibnamefont{and}
  \bibinfo{author}{\bibfnamefont{F.}~\bibnamefont{Hautmann}},
  \bibinfo{journal}{Phys. Lett.} \textbf{\bibinfo{volume}{B472}},
  \bibinfo{pages}{129} (\bibinfo{year}{2000}), \eprint{hep-ph/9908467}.

\bibitem[{\citenamefont{Manohar and Stewart}(2007)}]{zerobin}
\bibinfo{author}{\bibfnamefont{A.~V.} \bibnamefont{Manohar}} \bibnamefont{and}
  \bibinfo{author}{\bibfnamefont{I.~W.} \bibnamefont{Stewart}},
  \bibinfo{journal}{Phys. Rev.} \textbf{\bibinfo{volume}{D76}},
  \bibinfo{pages}{074002} (\bibinfo{year}{2007}), \eprint{hep-ph/0605001}.

\bibitem[{\citenamefont{Lee and Sterman}(2007)}]{leesterman}
\bibinfo{author}{\bibfnamefont{C.}~\bibnamefont{Lee}} \bibnamefont{and}
  \bibinfo{author}{\bibfnamefont{G.}~\bibnamefont{Sterman}},
  \bibinfo{journal}{Phys. Rev.} \textbf{\bibinfo{volume}{D75}},
  \bibinfo{pages}{014022} (\bibinfo{year}{2007}).

\bibitem[{\citenamefont{Idilbi and Mehen}(2007{\natexlab{a}})}]{idilbimehen1}
\bibinfo{author}{\bibfnamefont{A.}~\bibnamefont{Idilbi}} \bibnamefont{and}
  \bibinfo{author}{\bibfnamefont{T.}~\bibnamefont{Mehen}},
  \bibinfo{journal}{Phys. Rev.} \textbf{\bibinfo{volume}{D75}},
  \bibinfo{pages}{114017} (\bibinfo{year}{2007}{\natexlab{a}}),
  \eprint{hep-ph/0702022}.

\bibitem[{\citenamefont{Idilbi and Mehen}(2007{\natexlab{b}})}]{idilbimehen2}
\bibinfo{author}{\bibfnamefont{A.}~\bibnamefont{Idilbi}} \bibnamefont{and}
  \bibinfo{author}{\bibfnamefont{T.}~\bibnamefont{Mehen}},
  \bibinfo{journal}{Phys. Rev.} \textbf{\bibinfo{volume}{D76}},
  \bibinfo{pages}{094015} (\bibinfo{year}{2007}{\natexlab{b}}),
  \eprint{0707.1101}.

\bibitem[{\citenamefont{Chiu et~al.}(2009{\natexlab{b}})\citenamefont{Chiu,
  Fuhrer, Hoang, Kelley, and Manohar}}]{Delta}
\bibinfo{author}{\bibfnamefont{J.-Y.} \bibnamefont{Chiu}},
  \bibinfo{author}{\bibfnamefont{A.}~\bibnamefont{Fuhrer}},
  \bibinfo{author}{\bibfnamefont{A.~H.} \bibnamefont{Hoang}},
  \bibinfo{author}{\bibfnamefont{R.}~\bibnamefont{Kelley}}, \bibnamefont{and}
  \bibinfo{author}{\bibfnamefont{A.~V.} \bibnamefont{Manohar}},
  \bibinfo{journal}{Phys. Rev.} \textbf{\bibinfo{volume}{D79}},
  \bibinfo{pages}{053007} (\bibinfo{year}{2009}{\natexlab{b}}),
  \eprint{0901.1332}.

\bibitem[{\citenamefont{Chiu et~al.}(2009{\natexlab{c}})\citenamefont{Chiu,
  Fuhrer, Hoang, Kelley, and Manohar}}]{Chiu:2009yz}
\bibinfo{author}{\bibfnamefont{J.-y.} \bibnamefont{Chiu}},
  \bibinfo{author}{\bibfnamefont{A.}~\bibnamefont{Fuhrer}},
  \bibinfo{author}{\bibfnamefont{A.~H.} \bibnamefont{Hoang}},
  \bibinfo{author}{\bibfnamefont{R.}~\bibnamefont{Kelley}}, \bibnamefont{and}
  \bibinfo{author}{\bibfnamefont{A.~V.} \bibnamefont{Manohar}}
  (\bibinfo{year}{2009}{\natexlab{c}}), \eprint{0905.1141}.

\bibitem[{\citenamefont{Manohar}(2006)}]{messenger}
\bibinfo{author}{\bibfnamefont{A.~V.} \bibnamefont{Manohar}},
  \bibinfo{journal}{Phys. Lett.} \textbf{\bibinfo{volume}{B633}},
  \bibinfo{pages}{729} (\bibinfo{year}{2006}).

\bibitem[{\citenamefont{Bauer et~al.}(2002{\natexlab{b}})\citenamefont{Bauer,
  Fleming, Pirjol, Rothstein, and Stewart}}]{HardScattering}
\bibinfo{author}{\bibfnamefont{C.~W.} \bibnamefont{Bauer}},
  \bibinfo{author}{\bibfnamefont{S.}~\bibnamefont{Fleming}},
  \bibinfo{author}{\bibfnamefont{D.}~\bibnamefont{Pirjol}},
  \bibinfo{author}{\bibfnamefont{I.~Z.} \bibnamefont{Rothstein}},
  \bibnamefont{and} \bibinfo{author}{\bibfnamefont{I.~W.}
  \bibnamefont{Stewart}}, \bibinfo{journal}{Phys. Rev.}
  \textbf{\bibinfo{volume}{D66}}, \bibinfo{pages}{014017}
  (\bibinfo{year}{2002}{\natexlab{b}}).

\bibitem[{\citenamefont{Bohm et~al.}(2001)\citenamefont{Bohm, Denner, and
  Joos}}]{bohmbook}
\bibinfo{author}{\bibfnamefont{M.}~\bibnamefont{Bohm}},
  \bibinfo{author}{\bibfnamefont{A.}~\bibnamefont{Denner}}, \bibnamefont{and}
  \bibinfo{author}{\bibfnamefont{H.}~\bibnamefont{Joos}},
  \emph{\bibinfo{title}{{Gauge theories of the strong and electroweak
  interaction}}} (\bibinfo{publisher}{Teubner}, \bibinfo{address}{Stuttgart,
  Germany}, \bibinfo{year}{2001}).

\bibitem[{\citenamefont{Bardin and Passarino}(1999)}]{bardin}
\bibinfo{author}{\bibfnamefont{D.}~\bibnamefont{Bardin}} \bibnamefont{and}
  \bibinfo{author}{\bibfnamefont{G.}~\bibnamefont{Passarino}},
  \emph{\bibinfo{title}{The Standard Model in the Making}}
  (\bibinfo{publisher}{Oxford University Press}, \bibinfo{address}{Oxford},
  \bibinfo{year}{1999}).

\bibitem[{\citenamefont{Fleischer and Jegerlehner}(1981)}]{fleischer}
\bibinfo{author}{\bibfnamefont{J.}~\bibnamefont{Fleischer}} \bibnamefont{and}
  \bibinfo{author}{\bibfnamefont{F.}~\bibnamefont{Jegerlehner}},
  \bibinfo{journal}{Phys. Rev.} \textbf{\bibinfo{volume}{D23}},
  \bibinfo{pages}{2001} (\bibinfo{year}{1981}).

\bibitem[{\citenamefont{Bohm et~al.}(1986)\citenamefont{Bohm, Spiesberger, and
  Hollik}}]{hollik}
\bibinfo{author}{\bibfnamefont{M.}~\bibnamefont{Bohm}},
  \bibinfo{author}{\bibfnamefont{H.}~\bibnamefont{Spiesberger}},
  \bibnamefont{and} \bibinfo{author}{\bibfnamefont{W.}~\bibnamefont{Hollik}},
  \bibinfo{journal}{Fortsch. Phys.} \textbf{\bibinfo{volume}{34}},
  \bibinfo{pages}{687} (\bibinfo{year}{1986}).

\bibitem[{\citenamefont{Manohar}(1997)}]{amhqet}
\bibinfo{author}{\bibfnamefont{A.~V.} \bibnamefont{Manohar}},
  \bibinfo{journal}{Phys. Rev.} \textbf{\bibinfo{volume}{D56}},
  \bibinfo{pages}{230} (\bibinfo{year}{1997}), \eprint{hep-ph/9701294}.

\bibitem[{\citenamefont{Manohar}(1996)}]{ameft}
\bibinfo{author}{\bibfnamefont{A.~V.} \bibnamefont{Manohar}}
  (\bibinfo{year}{1996}), \eprint{hep-ph/9606222}.

\bibitem[{\citenamefont{Fleming
  et~al.}(2008{\natexlab{a}})\citenamefont{Fleming, Hoang, Mantry, and
  Stewart}}]{top1}
\bibinfo{author}{\bibfnamefont{S.}~\bibnamefont{Fleming}},
  \bibinfo{author}{\bibfnamefont{A.~H.} \bibnamefont{Hoang}},
  \bibinfo{author}{\bibfnamefont{S.}~\bibnamefont{Mantry}}, \bibnamefont{and}
  \bibinfo{author}{\bibfnamefont{I.~W.} \bibnamefont{Stewart}},
  \bibinfo{journal}{Phys. Rev.} \textbf{\bibinfo{volume}{D77}},
  \bibinfo{pages}{074010} (\bibinfo{year}{2008}{\natexlab{a}}),
  \eprint{hep-ph/0703207}.

\bibitem[{\citenamefont{Fleming
  et~al.}(2008{\natexlab{b}})\citenamefont{Fleming, Hoang, Mantry, and
  Stewart}}]{top2}
\bibinfo{author}{\bibfnamefont{S.}~\bibnamefont{Fleming}},
  \bibinfo{author}{\bibfnamefont{A.~H.} \bibnamefont{Hoang}},
  \bibinfo{author}{\bibfnamefont{S.}~\bibnamefont{Mantry}}, \bibnamefont{and}
  \bibinfo{author}{\bibfnamefont{I.~W.} \bibnamefont{Stewart}},
  \bibinfo{journal}{Phys. Rev.} \textbf{\bibinfo{volume}{D77}},
  \bibinfo{pages}{114003} (\bibinfo{year}{2008}{\natexlab{b}}),
  \eprint{0711.2079}.

\bibitem[{\citenamefont{Cornwall et~al.}(1974)\citenamefont{Cornwall, Levin,
  and Tiktopoulos}}]{cornwall}
\bibinfo{author}{\bibfnamefont{J.~M.} \bibnamefont{Cornwall}},
  \bibinfo{author}{\bibfnamefont{D.~N.} \bibnamefont{Levin}}, \bibnamefont{and}
  \bibinfo{author}{\bibfnamefont{G.}~\bibnamefont{Tiktopoulos}},
  \bibinfo{journal}{Phys. Rev.} \textbf{\bibinfo{volume}{D10}},
  \bibinfo{pages}{1145} (\bibinfo{year}{1974}).

\bibitem[{\citenamefont{Vayonakis}(1976)}]{vayonakis}
\bibinfo{author}{\bibfnamefont{C.~E.} \bibnamefont{Vayonakis}},
  \bibinfo{journal}{Nuovo Cim. Lett.} \textbf{\bibinfo{volume}{17}},
  \bibinfo{pages}{383} (\bibinfo{year}{1976}).

\bibitem[{\citenamefont{Lee et~al.}(1977)\citenamefont{Lee, Quigg, and
  Thacker}}]{leequiggthacker}
\bibinfo{author}{\bibfnamefont{B.~W.} \bibnamefont{Lee}},
  \bibinfo{author}{\bibfnamefont{C.}~\bibnamefont{Quigg}}, \bibnamefont{and}
  \bibinfo{author}{\bibfnamefont{H.~B.} \bibnamefont{Thacker}},
  \bibinfo{journal}{Phys. Rev.} \textbf{\bibinfo{volume}{D16}},
  \bibinfo{pages}{1519} (\bibinfo{year}{1977}).

\bibitem[{\citenamefont{Chanowitz and Gaillard}(1985)}]{chanowitz}
\bibinfo{author}{\bibfnamefont{M.~S.} \bibnamefont{Chanowitz}}
  \bibnamefont{and} \bibinfo{author}{\bibfnamefont{M.~K.}
  \bibnamefont{Gaillard}}, \bibinfo{journal}{Nucl. Phys.}
  \textbf{\bibinfo{volume}{B261}}, \bibinfo{pages}{379} (\bibinfo{year}{1985}).

\bibitem[{\citenamefont{Gounaris et~al.}(1986)\citenamefont{Gounaris, Kogerler,
  and Neufeld}}]{gounaris}
\bibinfo{author}{\bibfnamefont{G.~J.} \bibnamefont{Gounaris}},
  \bibinfo{author}{\bibfnamefont{R.}~\bibnamefont{Kogerler}}, \bibnamefont{and}
  \bibinfo{author}{\bibfnamefont{H.}~\bibnamefont{Neufeld}},
  \bibinfo{journal}{Phys. Rev.} \textbf{\bibinfo{volume}{D34}},
  \bibinfo{pages}{3257} (\bibinfo{year}{1986}).

\bibitem[{\citenamefont{Bagger and Schmidt}(1990)}]{bagger}
\bibinfo{author}{\bibfnamefont{J.}~\bibnamefont{Bagger}} \bibnamefont{and}
  \bibinfo{author}{\bibfnamefont{C.}~\bibnamefont{Schmidt}},
  \bibinfo{journal}{Phys. Rev.} \textbf{\bibinfo{volume}{D41}},
  \bibinfo{pages}{264} (\bibinfo{year}{1990}).

\bibitem[{\citenamefont{Yao and Yuan}(1988)}]{yaoyuan}
\bibinfo{author}{\bibfnamefont{Y.-P.} \bibnamefont{Yao}} \bibnamefont{and}
  \bibinfo{author}{\bibfnamefont{C.~P.} \bibnamefont{Yuan}},
  \bibinfo{journal}{Phys. Rev.} \textbf{\bibinfo{volume}{D38}},
  \bibinfo{pages}{2237} (\bibinfo{year}{1988}).

\bibitem[{\citenamefont{Denner et~al.}(1995)\citenamefont{Denner, Weiglein, and
  Dittmaier}}]{denner}
\bibinfo{author}{\bibfnamefont{A.}~\bibnamefont{Denner}},
  \bibinfo{author}{\bibfnamefont{G.}~\bibnamefont{Weiglein}}, \bibnamefont{and}
  \bibinfo{author}{\bibfnamefont{S.}~\bibnamefont{Dittmaier}},
  \bibinfo{journal}{Nucl. Phys.} \textbf{\bibinfo{volume}{B440}},
  \bibinfo{pages}{95} (\bibinfo{year}{1995}), \eprint{hep-ph/9410338}.

\bibitem[{\citenamefont{Moch et~al.}(2005)\citenamefont{Moch, Vermaseren, and
  Vogt}}]{moch:ff}
\bibinfo{author}{\bibfnamefont{S.}~\bibnamefont{Moch}},
  \bibinfo{author}{\bibfnamefont{J.~A.~M.} \bibnamefont{Vermaseren}},
  \bibnamefont{and} \bibinfo{author}{\bibfnamefont{A.}~\bibnamefont{Vogt}},
  \bibinfo{journal}{Phys. Lett.} \textbf{\bibinfo{volume}{B625}},
  \bibinfo{pages}{245} (\bibinfo{year}{2005}), \eprint{hep-ph/0508055}.

\bibitem[{\citenamefont{Mert~Aybat
  et~al.}(2006{\natexlab{a}})\citenamefont{Mert~Aybat, Dixon, and
  Sterman}}]{aybat1}
\bibinfo{author}{\bibfnamefont{S.}~\bibnamefont{Mert~Aybat}},
  \bibinfo{author}{\bibfnamefont{L.~J.} \bibnamefont{Dixon}}, \bibnamefont{and}
  \bibinfo{author}{\bibfnamefont{G.}~\bibnamefont{Sterman}},
  \bibinfo{journal}{Phys. Rev. Lett.} \textbf{\bibinfo{volume}{97}},
  \bibinfo{pages}{072001} (\bibinfo{year}{2006}{\natexlab{a}}),
  \eprint{hep-ph/0606254}.

\bibitem[{\citenamefont{Mert~Aybat
  et~al.}(2006{\natexlab{b}})\citenamefont{Mert~Aybat, Dixon, and
  Sterman}}]{aybat2}
\bibinfo{author}{\bibfnamefont{S.}~\bibnamefont{Mert~Aybat}},
  \bibinfo{author}{\bibfnamefont{L.~J.} \bibnamefont{Dixon}}, \bibnamefont{and}
  \bibinfo{author}{\bibfnamefont{G.}~\bibnamefont{Sterman}},
  \bibinfo{journal}{Phys. Rev.} \textbf{\bibinfo{volume}{D74}},
  \bibinfo{pages}{074004} (\bibinfo{year}{2006}{\natexlab{b}}),
  \eprint{hep-ph/0607309}.

\bibitem[{\citenamefont{Catani and Seymour}(1996)}]{catani}
\bibinfo{author}{\bibfnamefont{S.}~\bibnamefont{Catani}} \bibnamefont{and}
  \bibinfo{author}{\bibfnamefont{M.~H.} \bibnamefont{Seymour}},
  \bibinfo{journal}{Phys. Lett.} \textbf{\bibinfo{volume}{B378}},
  \bibinfo{pages}{287} (\bibinfo{year}{1996}), \eprint{hep-ph/9602277}.

\bibitem[{\citenamefont{Bauer and Manohar}(2004)}]{Bauer:2003pi}
\bibinfo{author}{\bibfnamefont{C.~W.} \bibnamefont{Bauer}} \bibnamefont{and}
  \bibinfo{author}{\bibfnamefont{A.~V.} \bibnamefont{Manohar}},
  \bibinfo{journal}{Phys. Rev.} \textbf{\bibinfo{volume}{D70}},
  \bibinfo{pages}{034024} (\bibinfo{year}{2004}), \eprint{hep-ph/0312109}.

\end{thebibliography}

\end{document}